\iffalse Before sending further, pass this check list:
  [ ] find and resolve all "??" and "\ifnote"
  [ ] Spell-check
  [ ] order of using/defining abbreviations
  [ ] citation order (use RefTest)
  [ ] figure order & reference order
  [ ] Check LaTeX  output files (*.log) for warnings
  [ ] Check BibTeX output (screen) for warnings
  [ ] update PACS
  [ ] decide on color figures
  [ ] Re-read paper in the morning

After completing this list, if you made at least one correction,
re-do this check-list from the beginning, until no corrections will
be done.\fi

\documentclass[amsmath,amssymb,prb,preprintnumbers,showpacs,superscriptaddress,twocolumn]{revtex4}

\usepackage{graphicx}
\usepackage{dcolumn}    
\usepackage{bm}         
\usepackage[usenames]{color} 
\usepackage{ulem}       
\usepackage[colorlinks=true, linkcolor=black, citecolor=black, urlcolor=black]{hyperref}		 
\usepackage{nicefrac} 				

\newcommand{\gapprox}{{\scriptscriptstyle\stackrel{>}{\sim}}}
\newcommand{\lapprox}{{\scriptscriptstyle\stackrel{<}{\sim}}}

\newif\ifgraph
\newif\ifcom
\newif\ifdel

\graphtrue                   %

\comtrue               %

\deltrue               %

\begin{document}

\title{Properties of the electron-doped infinite-layer superconductor Sr$_{1-x}$La$_{x}$CuO$_{2}$ epitaxially grown by pulsed laser deposition}

\author{J.~Tomaschko}
\affiliation{%
  Physikalisches Institut and Center for Collective Quantum Phenomena in LISA$^+$,
  Universit\"{a}t T\"{u}bingen,
  Auf der Morgenstelle 14,
  72076 T\"{u}bingen, Germany
}
\author{V.~Leca}
\email{leca@uni-tuebingen.de}
\affiliation{%
  Physikalisches Institut and Center for Collective Quantum Phenomena in LISA$^+$,
  Universit\"{a}t T\"{u}bingen,
  Auf der Morgenstelle 14,
  72076 T\"{u}bingen, Germany
}
\affiliation{%
  National Institute for Research and Development in Microtechnologies, Erou Iancu Nicolae Str. 126~A, 077190, Bucharest, Romania
}
\affiliation{%
	Faculty of Applied Chemistry and Materials Science, University Politehnica of Bucharest,
Gheorghe Polizu Str. 1-7, 011061, Bucharest, Romania
}
\author{T.~Selistrovski}
\affiliation{%
  Physikalisches Institut and Center for Collective Quantum Phenomena in LISA$^+$,
  Universit\"{a}t T\"{u}bingen,
  Auf der Morgenstelle 14,
  72076 T\"{u}bingen, Germany
}
\author{S.~Diebold}
\affiliation{%
  Physikalisches Institut and Center for Collective Quantum Phenomena in LISA$^+$,
  Universit\"{a}t T\"{u}bingen,
  Auf der Morgenstelle 14,
  72076 T\"{u}bingen, Germany
}
\author{J.~Jochum}
\affiliation{%
  Physikalisches Institut and Center for Collective Quantum Phenomena in LISA$^+$,
  Universit\"{a}t T\"{u}bingen,
  Auf der Morgenstelle 14,
  72076 T\"{u}bingen, Germany
}
\author{R.~Kleiner}
\author{D.~Koelle}
\affiliation{%
  Physikalisches Institut and Center for Collective Quantum Phenomena in LISA$^+$,
  Universit\"{a}t T\"{u}bingen,
  Auf der Morgenstelle 14,
  72076 T\"{u}bingen, Germany
}

\date{\today}

\begin{abstract}
Thin films of the electron-doped infinite-layer cuprate superconductor Sr$_{1-x}$La$_x$CuO$_2$ (SLCO) with doping $x \approx 0.15$ were grown by means of pulsed laser deposition.
(001)-oriented KTaO$_3$ and SrTiO$_3$ single crystals were used as substrates.
In case of SrTiO$_3$, a BaTiO$_3$ thin film was deposited prior to SLCO, acting as buffer layer providing tensile strain to the SLCO film.
To induce superconductivity, the as-grown films were annealed under reducing conditions, which will be described in detail.
The films were characterized by reflection high-energy electron diffraction, atomic force microscopy, x-ray diffraction, Rutherford backscattering spectroscopy, and electric transport measurements at temperatures down to $T = 4.2\,$K.
We discuss in detail the influence of different process parameters on the final film properties.
\end{abstract}


\maketitle

\section{Introduction}
\label{sec:Introduction}

Cuprates exhibiting electron doping form a minor group among high-transition temperature (high-$T_c$) cuprate superconductors.
In fact, only two families of electron-doped cuprate superconductors are known.
These are the $T'$-compounds\cite{Tokura89, Takagi89} $Ln^\mathrm{III}_{2-x}$Ce$_x$CuO$_4$ ($Ln^\mathrm{III} = \mathrm{La, Pr, Nd, Sm, Eu, Gd}$) with maximum $T_c \approx 30\,$K\cite{Yamada94,Naito00} and the infinite-layer (IL) compounds\cite{Siegrist88,Smith91,Er91} Sr$_{1-x}Ln^\mathrm{III}_x$CuO$_2$ ($Ln^\mathrm{III} = \mathrm{La, Pr, Nd, Sm, Gd}$) with maximum $T_c \approx 43\,$K\cite{Smith91,Er92,Jorgensen93,Ikeda93}.
Formally, the IL crystal is a member of the $A^\mathrm{III}_{2}A^\mathrm{II}_2$Ca$_{n-1}$Cu$_n$O$_{2n+4}$ high-$T_c$ superconductors ($A^\mathrm{III}$ = Tl or Bi, and $A^\mathrm{II}$ = Ca, Sr or Ba), where $n = \infty$\cite{Torardi88,Siegrist88}.
The IL crystal is thus formed by alternating stacks of copper oxide planes (CuO$_2$) and alkaline earth metal planes ($A^\mathrm{II}$) along the $c$-axis direction, forming an $A^\mathrm{II}$CuO$_2$ crystal.
To induce electron doping, the divalent alkaline earth metal $A^\mathrm{II}$ is substituted partially by a trivalent lanthanide $Ln^\mathrm{III}$ to form $A^\mathrm{II}_{1-x}Ln^\mathrm{III}_x$CuO$_2$, or as in our case Sr$_{1-x}$La$_x$CuO$_2$ (SLCO)\cite{Er92}.
Common features of cuprate superconductors, such as apical oxygen or charge reservoir blocks, are not present in the ideal IL structure\cite{Shaked95}.
As the crystal structure of the IL compounds is the most simple of all cuprate superconductors, they are often denoted as "parent structure" of cuprate superconductors\cite{Siegrist88}.
Due to their simplicity, they also provide a unique opportunity to explore the basic nature of high-$T_c$ superconductivity.
However, synthesizing high-quality IL samples is a challenging task, which explains that IL compounds have been examined less extensively than other superconducting cuprates.
No large IL single crystals have been synthesized so far, and only high-pressure synthesis ($\sim 1\,$GPa) of polycrystalline bulk material was successful\cite{Er91, Ikeda93, Jorgensen93}.
To overcome this problem, single crystalline thin films were grown, where the high-pressure IL phase is stabilized by epitaxy\cite{Li92,Terashima93}.
However, in the first attempts, SrTiO$_3$ (STO) with an in-plane lattice constant of $a_\mathrm{STO} = 3.905\,$\AA\, was most often used as substrate\cite{Niu92,Sugii92}, leading to compressively strained IL films (bulk $a_\mathrm{IL} \approx 3.95\,$\AA) with inferior superconducting properties.
It is known, that electron doping stretches the Cu$-$O bonds because antibonding $\sigma_{x^2-y^2}$ orbitals in the CuO$_2$ sheets are occupied\cite{Er91}.
Therefore, the idea was to enhance the electron-doping effect by epitaxial strain.
Different buffer layers, such as $T'$-compounds ($Ln^\mathrm{III}_2$CuO$_4$) or BaTiO$_3$ (BTO), with increased in-plane lattice constants were subsequently introduced to induce relaxed or tensile strained films\cite{Sugii93,Markert00,Markert03,Leca06,Leca08}.
Indeed, superconducting IL films were fabricated with this method, however still with reduced $T_c$.
Only one group\cite{Karimoto01} succeeded in synthesizing electron-doped IL films with $T_c$ close to the bulk value by means of molecular beam epitaxy (MBE).
The key was to choose KTaO$_3$ (KTO, with $a_\mathrm{KTO} = 3.988\,$\AA) as substrate, supplying tensile strain and making buffer layers redundant.
However, in a later work\cite{Karimoto04}, the same group showed that they could further increase $T_c$ when they chose another substrate, (110)-oriented DyScO$_3$ (DSO), with a slightly smaller in-plane lattice constant $\sim$$3.944\,$\AA, which fits better the lattice constant of bulk Sr$_{0.9}$La$_{0.1}$CuO$_2$ ($a_\mathrm{SLCO} \approx 3.949\,$\AA)\cite{Er92,Jorgensen93}.
Thus, growth of SLCO on DSO resulted in almost fully relaxed SLCO films with best transport properties reported, so far.
\\
Moreover, it was found that as-grown IL films contain excess oxygen forming O$^{2-}$ ions on interstitial sites in the $A^\mathrm{II}_{1-x}Ln^\mathrm{III}_{x}$-planes, which hampers superconductivity by localizing free charge carriers and by disturbing\cite{Jorgensen94} the crystal lattice.
Therefore, as for $T'$-compounds\cite{Naito97}, a vacuum annealing step was introduced to remove excess oxygen and to induce superconductivity.
Meanwhile, this reduction step is commonly used for synthesis of superconducting IL films grown by numerous techniques, such as sputtering\cite{Adachi92, Li09}, pulsed laser deposition (PLD)\cite{Leca06,Leca08} and MBE\cite{Karimoto01,Karimoto04}.
However, too strong reduction generates oxygen vacancies in the CuO$_2$ planes and destroys superconductivity\cite{Karimoto01,Li09,Bals03}.
A secondary phase can be formed if the oxygen vacancies arrange in an ordered structure, referred to in literature as the "long $c$-axis" phase or "infinite-layer-related" (IL-r) phase\cite{Adachi92,Zhou93,Mercey95,Leca03,Leca06}.
Its unit cell $2\sqrt{2}a_\mathrm{IL}\times 2\sqrt{2}a_\mathrm{IL}\times c_s$ is a superstructure of the IL unit cell, where $a_\mathrm{IL}$ is the in-plane lattice parameter of the IL structure and $c_s$ is the extended $c$-axis parameter of the superstructure (with $c_s\sim3.6\,${\AA} as compared to $c_\mathrm{IL}\sim3.4\,${\AA}).
Hence, the main challenge in synthesizing superconducting IL compounds is to simultaneously reduce the $A^\mathrm{II}_{1-x}Ln^\mathrm{III}_{x}$-planes without reducing the CuO$_2$-planes.
\\
In this work, we report on the fabrication of superconducting SLCO films (with maximum $T_c = 22\,$K) with doping $x \approx 0.15$ by means of PLD.
The fabrication process is described in detail.
KTO and STO single crystals were used as substrates.
Prior to deposition of SLCO on STO, BTO films were deposited, acting as buffer layers.
The growth mode, the evolution of the in-plane lattice parameter $a$, and the morphology of the films was monitored in-situ by high-pressure reflection high-energy electron diffraction (RHEED).
Atomic force microscopy (AFM) revealed very flat surfaces with asperities in the range of 1\,-\,3 unit cells (uc).
X-ray diffraction (XRD) was used to check the crystal quality and lattice constants of the films, revealing almost completely relaxed BTO buffer layers, moderately tensile strained SLCO films on BTO-buffered STO and highly tensile strained SLCO films on KTO.
The stoichiometry of SLCO films on BTO/STO was determined by Rutherford backscattering spectroscopy (RBS), showing that the films are slightly overdoped.
Current-voltage ($I(V)$) characteristics and resistivity $\rho$ vs temperature $T$ were examined by electric transport measurements at temperatures down to $T = 4.2\,$K.
Finally, we discuss the influence of different process parameters on the final film properties, such as excimer laser energy, target-to-substrate distance, deposition pressure, vacuum annealing time, and vacuum annealing temperature.

\section{Sample fabrication and experimental details}
\label{sec:fabrication}

For epitaxial growth of typically 20 to 25\,nm thick SLCO films, a polycrystalline target\cite{Chemco} with nominal doping $x = 0.125$ was used.
Both, (001)-oriented KTO\cite{Crystal} and (001)-oriented STO\cite{Crystec} single crystals ($5 \times 5 \times 1\,$mm$^3$) were used as substrates.
When STO was used as substrate, a typically 25 to 30\,nm thick BTO film was deposited prior to SLCO, acting as a buffer layer.
For that purpose, a stoichiometric, polycrystalline target\cite{Chemco} was used.
All films were grown in an ultra high vacuum system (base pressure $p_\mathrm{vac} \approx 10^{-6}$\,Pa) by means of PLD, using a KrF excimer laser ($\lambda = 248\,$nm) at a repetition rate of $2\,$Hz.
The excimer laser energy was set to $E_L = ($110\,-\,130)\,mJ and the laser spot size on the target was $A_L \approx 2\,$mm$^2$.
The growth mode and the number of deposited monolayers were monitored in-situ by high-pressure RHEED.
The sample temperature was checked with a band radiation pyrometer (1.45 to 1.80\,$\mu\mathrm{m}$) while the sample was heated with an infrared diode laser ($\lambda = 808\,$nm) irradiating the backside of the sample holder.
Right after SLCO film deposition, 10 to 20\,nm thick gold pads were evaporated or magnetron sputtered ex-situ through shadow masks, to ensure low-resistive ohmic contacts.
For electric transport measurements, we performed $\rho(T)$ measurements on unpatterned films in a 4-point van der Pauw geometry with typical bias currents $I = ($10\,-\,100)\,$\mu$A.
All measured values for $T_c$ given in the text refer to the midpoint of the resistive transition.
For measurements of $I(V)$ characteristics, the films were patterned into $40\,\mu$m wide bridges (with 20 or $40\,\mu$m voltage pad separation) by photolithography and argon ion milling.
$I(V)$ and $\rho(T)$ curves were recorded in a magnetically and radio frequency shielded setup using feed lines with high-frequency noise filters.
The crystal structure of the films was characterized ex-situ by XRD equipped with a Cu cathode and monochromator.
The morphology was checked by AFM in contact mode.
RBS was performed at a 3\,MeV Van-de-Graaff accelerator to determine the stoichiometry of the films\cite{Rosenau,Diebold10}.
The accelerator can be used at energies between 0.7 and 3.7\,MeV with a beam stability of $\sim 2\,$keV.
The pressure is (0.5\,-\,1$) \times 10^{-4}\,$Pa and the acceleration distance is 30\,m.
RBS was performed with $\alpha$-particles at a fixed angle of 165$\,^\circ$.
The energy resolution of the detector is 20\,keV.

\subsection{Deposition of BTO buffer layers on STO}
\label{subsec:btosto}

For as-received STO substrates, we established an in-situ annealing process prior to BTO deposition to ensure the formation of a smooth STO surface suitable for epitaxial growth of BTO.
For that purpose, the STO substrates were heated in oxygen at a pressure of $p_\mathrm{O_2} = 10\,$Pa to $T = 1000\,^\circ$C with a rate of $\delta T / \delta t = 40\,^\circ\mathrm{C}/$min and kept there for $t = 5\,$min.
During this period, the RHEED pattern became more pronounced, exhibiting several Kikuchi lines and thin streaks and spots (cf.~Fig.~\ref{fig:btorheed}~(a)), typical for smoothing of the STO surface and desorption of surface contaminants.
This annealing step further helped to enhance the reproducibility of thermal coupling between substrate and sample holder which were fixed to each other with silver paste.
Moreover, if STO was heated in vacuum instead of oxygen, we observed slight intermediate streaks in the RHEED pattern (not shown here), which we explain by the formation of a superstructure, possibly an oxygen-deficient phase.
After STO annealing, the substrate was cooled down in $p_\mathrm{O_2} = 10\,$Pa oxygen to $T = 700\,^\circ$C with a rate of $\delta T / \delta t = -40\,^\circ\mathrm{C}/$min.
The power of the heating diode laser was locked during preablation and ablation, leading to a sample temperature typically in the range of $T = ($670\,-\,700)\,$^\circ$C.
The energy of the excimer laser was set to $E_L = 110\,$mJ at a target-to-substrate distance $d_\mathrm{TS} = 65\,$mm.
Before ablation, the target was preablated with 500 laser pulses to get a clean and stoichiometric surface.
Then, 900 pulses at a repetition rate of $f_L = 2\,$Hz were used for deposition of the BTO film.
As the morphology of BTO is of great importance for the properties of the SLCO films, the as-deposited BTO films had to be annealed.
Therefore, the sample was heated up with $\delta T / \delta t = 40\,^\circ$C/min in deposition pressure $p_\mathrm{O_2} = 10\,$Pa to $T = 900\,^\circ$C and annealed for $t_a = 15\,$min.
We found that heating and anneling in oxygen is essential because bare vacuum annealing led to the formation of a superstructure, probably an oxygen deficient phase, as observed by the formation of intermediate streaks in the RHEED pattern (not shown here).
After oxygen annealing, the oxygen was turned off and the samples were annealed additionally for $t_a = 30\,$min in vacuum at $T = 900\,^\circ$C to remove possible excess oxygen from BTO, which could diffuse into the SLCO and hamper the reduction of the latter.
An analysis of the BTO annealing procedure will be given in sec.~\ref{subsubsec:btorheedafm}. 
After oxygen and vacuum annealing, the BTO films were cooled down in vacuum ($p_\mathrm{vac} \,\lapprox\, 10^{-5}\,$Pa) at a rate of $\delta T / \delta t = -20\,^\circ$C/min to the deposition temperature of SLCO.

\subsection{Deposition of SLCO on BTO buffered STO}
\label{subsec:slcobtosto}

The SLCO films were deposited in-situ after BTO had been deposited on STO as described in sec.~\ref{subsec:btosto}.
At $T = 550\,^\circ$C the pressure was increased to $p_\mathrm{O_2} = 20\,$Pa.
The power of the heating diode laser was locked during preablation and ablation, resulting in a temperature of $T = (575 \pm 10)\,^\circ$C.
To minimize reoxidation of the previously vacuum annealed BTO film, the preablation of SLCO was started immediately after turning on the oxygen flow.
The excimer laser was set at $f_L = 2\,$Hz at an energy of $E_L = 130\,$mJ and the target-to-substrate distance to $d_\mathrm{TS} = 60\,$mm.
500 pulses were preablated and (600\,-\,850) pulses were ablated.
To remove excess oxygen incorporated during SLCO growth, the oxygen flow was turned off immediately after deposition and the sample was annealed in vacuum for typically $t_a = (20 \pm 5)\,$min.
During vacuum annealing, the temperature increased to $T = (600 \pm 10)\,^\circ$C and the pressure decreased to $p_\mathrm{vac} \approx 10^{-5}\,$Pa.
The streaky RHEED pattern got more pronounced but slight intermediate streaks, formed during deposition probably due to excess oxygen, did not vanish completely (cf.~Fig.~\ref{fig:slcobtorheedafm}~(a)).
After vacuum annealing, the heating diode laser was turned off and the sample cooled down in vacuum to room temperature within $t \approx 1\,$h.

\subsection{Deposition of SLCO on KTO}
\label{subsec:slcokto}

Contrary to STO, KTO substrates were not annealed in vacuum at high $T$ prior to deposition, because it is known that such annealing leads to the formation of a rough surface, probably due to the formation of reduced forms of tantalum oxide on the surface\cite{Leca03}. 
Thus, KTO was heated up in vacuum ($p_\mathrm{vac} \approx 10^{-5}\,$Pa) to $T = 550\,^\circ$C at a heating rate of $\delta T / \delta t = 10\,^\circ$C/min.
Then, the pressure inside the PLD chamber was increased to $p_\mathrm{O_2} = 20\,$Pa and the power of the heating diode laser was locked, leading to a substrate temperature of $T = (580 \pm 10)\,^\circ$C.
The target-to-substrate distance was adjusted to $d_\mathrm{TS} = 60\,$mm and the energy of the excimer laser to $E_L = 130\,$mJ at a repetition rate of $f_L = 2\,$Hz.
500 and 850 pulses were preablated and ablated, respectively.
To remove excess oxygen, the oxygen flow was turned off immediately after deposition and the sample was vacuum annealed for typically $t_a = (10 \pm 5)\,$min.
During annealing, the pressure decreased to $p_\mathrm{vac} \approx 10^{-5}\,$Pa and the temperature increased to $T = (605 \pm 5)\,^\circ$C.
Finally, the heating diode laser was turned off and the sample cooled down to room temperature.

\section{Characterization of thin films}
\label{sec:charcterization}

In this section, typical properties of the BTO and SLCO films are presented.
For simplicity, only representative measurements are shown and discussed.
The influence of different process parameters on the final film properties will be discussed in sec.~\ref{sec:discussion}.

\subsection{Characterization of BTO on STO}
\label{subsec:charbto}

\subsubsection{RHEED and atomic force microscopy}
\label{subsubsec:btorheedafm}

\begin{figure*}[tb]
\centering
\ifgraph\includegraphics[width=0.95\textwidth]{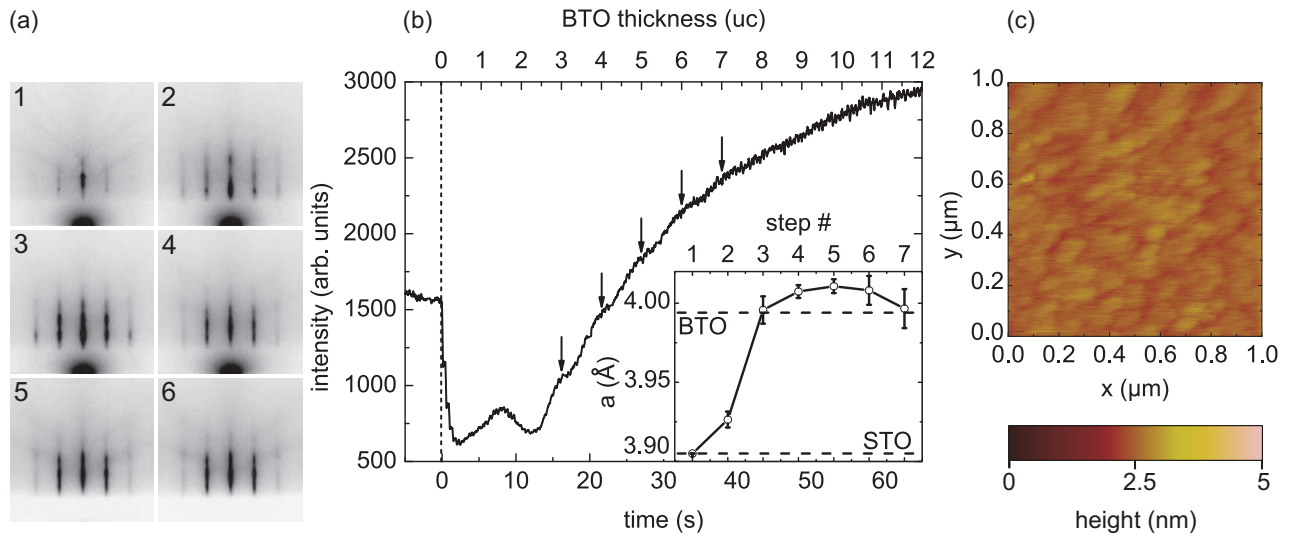}\fi
\caption{(Color online)
(a) RHEED patterns during preparation of BTO. 1: STO substrate before deposition, 2: BTO after growth of $\sim 2$ unit cells, 3: end of BTO deposition, 4: begin of BTO oxygen annealing, 5: begin of BTO vacuum annealing, 6: end of BTO vacuum annealing. (b) Main graph shows intensity oscillations of the RHEED specular spot during growth of BTO (arrows). Inset shows the evolution of the in-plane lattice constant $a_\mathrm{BTO}$ during preparation of BTO as derived from RHEED patterns illustrated in (a). The value of step \#7 was determined after cooling the sample in vacuum. Horizontal lines indicate the in-plane lattice constants of STO (literature value 3.905\,\AA) and BTO (determined by XRD, 3.994\,\AA). (c) AFM image of the BTO surface. The root mean square roughness is 0.15\,nm and the maximum step height is 0.8\,nm, corresponding to 2\,uc BTO.
}
\label{fig:btorheed}
\end{figure*}
%
During deposition, heating and annealing, the evolution of the in-plane lattice constant $a_\mathrm{BTO}$ and the morphology of the BTO films were analyzed by means of RHEED, as shown in Figs.~\ref{fig:btorheed}~(a) and (b).
By use of Bragg's law ($2 a \sin \theta_n = n \lambda$, where $\theta_n$ is the angle of the $n^\mathrm{th}$ order maximum and $\lambda$ the de Broglie wavelength of the diffracted electrons) $a_\mathrm{BTO}$ can be determined.
Small angle approximation yields $a_\mathrm{BTO}/a_\mathrm{STO} \approx \theta_{n,\mathrm{STO}} / \theta_{n,\mathrm{BTO}}$.
With the lattice parameter $a_\mathrm{STO} = 3.905\,$\AA\cite{CRC71} and with the angles $\theta_n$ extracted from RHEED patterns, we determined $a_\mathrm{BTO}$.
The inset of Fig.~\ref{fig:btorheed}~(b) shows the mean value of $a_\mathrm{BTO}$ determined during growth of 10 comparable BTO films.
Error bars denote the standard deviation.
We found, that the first unit cells of BTO were highly compressively strained because after deposition of $\sim 2\,$uc BTO (cf.~step \#2 in Fig.~\ref{fig:btorheed}), the lattice constant was determined as $a_\mathrm{BTO} \approx 3.92\,$\AA, which is close to $a_\mathrm{STO}$. 
Yet, due to the large lattice mismatch of $2.2\,$\% between bulk BTO with $a_\mathrm{BTO} = (3.992 \pm 0.002)\,$\AA\cite{Dungan52, Donohue58} and STO, the BTO films relaxed during growth, which was observed as an increase of $a_\mathrm{BTO}$ from its initial value to $\sim 3.99\,$\AA\, at the end of deposition (step \#3).
The relaxation of BTO was further directly visible in the RHEED pattern because three-dimensional (3D) signatures occured in the initially two-dimensional (2D) pattern during deposition (cf.~step \#2 and \#3).
This allows us to identify the growth mode of the BTO films as a mixture of 2D layer-by-layer and 3D island growth, which is known as Stranski-Krastanov growth\cite{Stranski39}.
Moreover, this observation is in accordance with the disappearance of intensity oscillations of the RHEED specular spot after $\sim 5$\,uc, as depicted in Fig.~\ref{fig:slcobtorheedafm}~(b).
During subsequent heating in oxygen (steps \#3 - 4) and annealing in oxygen (steps \#4 - 5), a smoothing of the BTO surface was observed because 3D admixtures diminished, leaving a mostly streaky, 2D pattern with Kikuchi lines.
Moreover, $a_\mathrm{BTO}$ increased to $\sim 4.00\,$\AA\, after oxygen annealing (step \#4).
Such an increase of $a$, usually going along with a decrease of $c$, is a common feature observed during oxidation of oxides\cite{Werner09}.
During vacuum annealing (steps \#5 - 6) and cooling (steps \#6 - 7) the BTO film was reduced, resulting in a decrease of $a_\mathrm{BTO}$ to its final value of $\sim 3.995\,$\AA.
As RHEED is a surface sensitive method, contrary to XRD, the final in-plane lattice constant $a_\mathrm{BTO} \approx 3.995\,$\AA\, indeed corresponds to the \textit{uppermost} unit cells, which are decisive for epitaxial growth of SLCO.
In conclusion, tensile strain can be provided to SLCO by use of BTO as buffer layer.
\\
The morphology of the BTO films was furthermore examined by AFM.
As shown in Fig.~\ref{fig:btorheed}~(c), AFM revealed a very smooth surface with a root mean square roughness $RMS \approx 0.15\,$nm and a maximum step height of $\sim 0.8\,\mathrm{nm}$.
This result corresponds to steps of 1 to 2\,uc and is in accordance with a smooth BTO surface as observed by RHEED.

\subsubsection{X-ray diffraction}
\label{subsubsec:btoxrd}

\begin{figure*}[tb]
\centering
\ifgraph\includegraphics[width=1.00\textwidth]{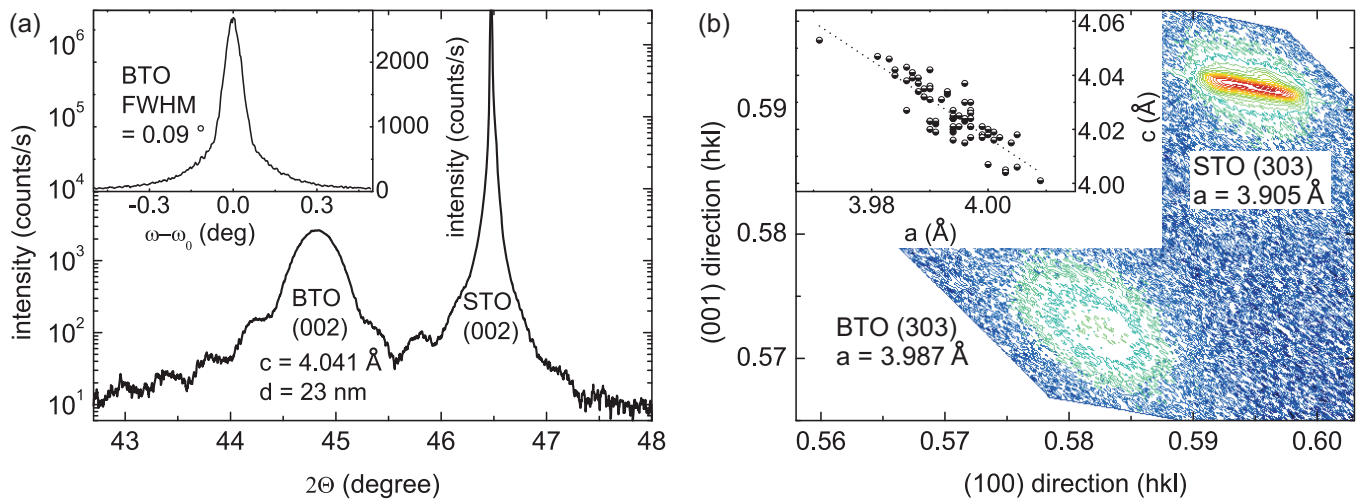}\fi
\caption{(Color online)
XRD data of BTO on STO. (a) Main graph shows the (002) region of a $\Theta - 2 \Theta$ scan and inset shows the rocking curve of the BTO (002) peak. (b) Main graph shows reciprocal space mapping of the (303) region of SLCO and BTO. Inset displays the linear dependence of the BTO lattice parameters $a_\mathrm{BTO}$ and $c_\mathrm{BTO}$. The dashed line is a linear fit.
}
\label{fig:btoxrd}
\end{figure*}
%
Figure \ref{fig:btoxrd} shows XRD data of a typical BTO film on STO.
The film is aligned along the $a$-/$b$-axis of STO, as found by $\Phi$ scans around the sample normal (not shown here).
The BTO (002) peak yields a $c$-axis constant of 4.041\,\AA.
Interference fringes around the (002) peak of BTO in Fig.~\ref{fig:btoxrd}~(a), known as Laue oscillations\cite{Warren69}, indicate high crystalline quality along the film normal and a flat surface.
Narrow rocking curves with typical full width at half maximum $\mathrm{FWHM} \approx 0.1\,^\circ$ (cf.~inset of Fig.~\ref{fig:btoxrd}~(a)) confirm the high crystalline quality of the films.
\\
The reciprocal space map in the main graph of Fig.~\ref{fig:btoxrd}~(b) yields an in-plane lattice constant of $a_\mathrm{BTO} = 3.989\,$\AA.
As XRD provides integral information on the crystal structure, the very sharp BTO (303) peak without any extension along the (100) direction shows that the major part of the film is relaxed.
This observation is in accordance with the RHEED results presented in sec.~\ref{subsubsec:btorheedafm}, which showed that the film was almost fully relaxed after annealing.
\\
The inset of Fig.~\ref{fig:btoxrd}~(b) displays the correlation of the lattice parameters $a_\mathrm{BTO}$ and $c_\mathrm{BTO}$ for $\sim 60$ BTO films prepared under varying conditions.
In particular, the deposition temperature was varied between $T = 650$ and 750\,K and some films were only annealed in oxygen, whereas others were additionally annealed in vacuum.
These variations most likely result in BTO films with varying oxygen content.
As commonly known, an increase of $a_\mathrm{BTO}$ leads to a decrease of $c_\mathrm{BTO}$, which is ascribed to an increasing degree of oxidation.
This assumption will be confirmed in sec.~\ref{subsec:procbto}.
We determined mean values of $a_\mathrm{BTO} = (3.994 \pm 0.006)\,$\AA\, and $c_\mathrm{BTO} = (4.026 \pm 0.011)\,$\AA, which are close to the bulk lattice constants $a_\mathrm{BTO} = (3.992 \pm 0.002)\,$\AA\, and $c_\mathrm{BTO} = (4.031 \pm 0.002)\,$\AA\cite{Dungan52, Donohue58}, confirming that the BTO films are relaxed.
Note, that the XRD value of $a_\mathrm{BTO}$ coincides with the \textit{surface} value determined by RHEED (cf.~sec.~\ref{subsubsec:btorheedafm}).
Thus, the surface layer is coherently relaxed with the bulk.
However, RHEED oscillations of the specular spot, observed at the beginning of deposition, yielded an extrapolated film thickness of $(79 \pm 4)\,$uc, i.e.~$t_\mathrm{BTO} = (32 \pm 2)\,$nm, whereas Laue oscillations of XRD $\Theta - 2\Theta$ scans revealed a film thickness typically 2-3\,nm smaller than the value determined by RHEED.
This points to a thin layer, which is non-coherently strained with the bulk.
We interpret this as a compressively strained BTO layer close to the interface with STO.
As known from literature, such an interface layer does exist and it comprises most of the dislocations\cite{Terai02}.
It is likely that such a strained interface layer does not contribute to Laue oscillations.
\\
To conclude with sec.~\ref{subsec:charbto}, we found that the BTO films have a flat and relaxed surface, which makes them suitable to be used as buffer layers for epitaxial growth of tensile strained SLCO films.

\subsection{Characterization of SLCO on BTO/STO}
\label{subsec:charslcobtosto}

\subsubsection{RHEED and atomic force microscopy}
\label{subsubsec:slcobtorheedafm}

\begin{figure*}[tb]
\centering
\ifgraph\includegraphics[width=0.95\textwidth]{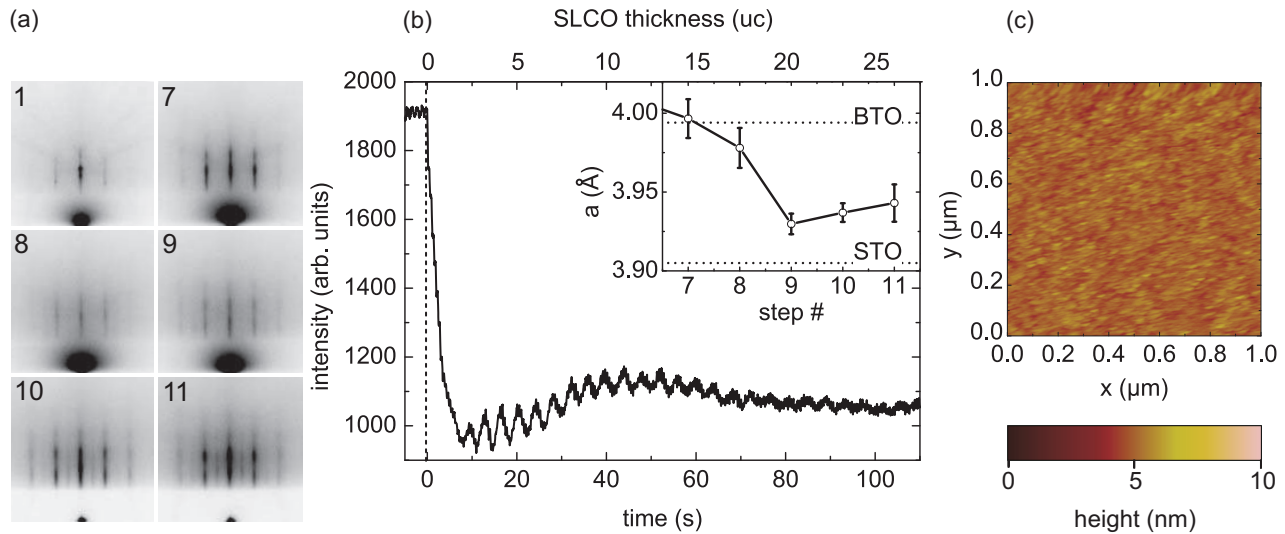}\fi
\caption{(Color online)
(a) RHEED patterns during preparation of SLCO on BTO/STO. 1: STO substrate before deposition, 7: BTO buffer layer before SLCO deposition, 8: SLCO after growth of $\sim 2$ unit cells, 9: end of SLCO deposition, 10: begin of SLCO vacuum annealing, 11: end of SLCO vacuum annealing. (b) Main graph shows intensity oscillations of the RHEED specular spot during growth of SLCO on BTO/STO. Deposition started at $t = 0$. Inset shows the evolution of the in-plane lattice constant $a_\mathrm{SLCO}$ during preparation of SLCO as derived from RHEED patterns illustrated in (a). Steps \#1 - \#7 are discussed in sec.~\ref{subsubsec:btorheedafm}. Horizontal lines indicate the in-plane lattice constants of STO (literature value 3.905\,\AA) and BTO (mean value determined by XRD, 3.994\,\AA). (c) AFM image of the SLCO surface. The root mean square roughness is 0.35\,nm and the maximum step height is 1.0\,nm, corresponding to 3\,uc SLCO.
}
\label{fig:slcobtorheedafm}
\end{figure*}
%
For SLCO films deposited on BTO/STO, typically $(20 \pm 10)$ intensity oscillations of the RHEED specular spot could be observed before they vanished (cf.~Fig.~\ref{fig:slcobtorheedafm}~(b)).
The RHEED oscillations revealed a growth rate of $(8.2 \pm 0.5)\,$pls/uc, yielding a film thickness of $t_\mathrm{SLCO} = (26 \pm 2)\,$nm for 600 pulses.
At the end of deposition, sharp and streaky RHEED patterns with some fade 3D dots as a result of increased surface disorder were observed as illustrated in Fig.~\ref{fig:slcobtorheedafm}~(a).
As in the case of BTO, this implies a Stranski-Krastanov growth mode.
However, weak intermediate streaks in the RHEED pattern could be observed during film growth which did not vanish during annealing (cf.~Fig.~\ref{fig:slcobtorheedafm}~(a)).
Note, that for identical settings those streaks did not appear when SLCO was deposited on KTO.
The possible origin of the streaks will be discussed in sec.~\ref{subsubsec:annealtime}, hinting at excess oxygen which could not be removed.
\\
The inset of Fig.~\ref{fig:slcobtorheedafm}~(b) illustrates the evolution of the in-plane lattice constant $a_\mathrm{SLCO}$ during deposition and annealing of SLCO, where step \#7 corresponds to step \#7 of Fig.~\ref{fig:btorheed}.
The data was acquired from 10 samples fabricated under similar conditions and the error bars represent the standard deviation.
As derived from step \#8, SLCO begins to relax immediately.
At the end of deposition, it has an in-plane lattice constant of $a_\mathrm{SLCO} \approx 3.93\,$\AA, which is remarkably small as compared to the bulk value of $3.967\,$\AA\, determined by XRD after vacuum annealing (cf.~sec.~\ref{subsubsec:slcobtoxrd}).
We explain this as follows.
During deposition, excess oxygen is incorporated in the SLCO film, leading to an elongated $c$-axis and a shortened $a$-axis.
During vacuum annealing, excess oxygen desorbs, leading to an increase of the $a$-axis, as observed by RHEED in steps \#10 - \#11.
However, the SLCO bulk value determined by XRD is not reached.
This discrepancy is supposably due to the fact that RHEED is a surface sensitive method, yielding different lattice parameters than found for the bulk because of different surface structure or composition.
Indeed, in a recent paper, we have proven the existence of a thin ($\sim 3\,$nm), oxygen deficient SLCO surface layer, which most probably forms during vacuum annealing.\cite{Tomaschko11}
Such a reduced SLCO surface layer explains well the small in-plane parameter $a_\mathrm{SLCO}$ observed by RHEED.
\\
Additionally to RHEED, we checked the morphology of the SLCO films with AFM.
A typical AFM image is shown in Fig.~\ref{fig:slcobtorheedafm}~(c).
The $RMS$ was determined as 0.35\,nm and the maximum step height as 1.0\,nm, corresponding to 3\,uc SLCO.
Therefore, the film is quite flat, confirming the RHEED results.

\subsubsection{X-ray diffraction}
\label{subsubsec:slcobtoxrd}

\begin{figure*}[tb]
\centering
\ifgraph\includegraphics[width=1.00\textwidth]{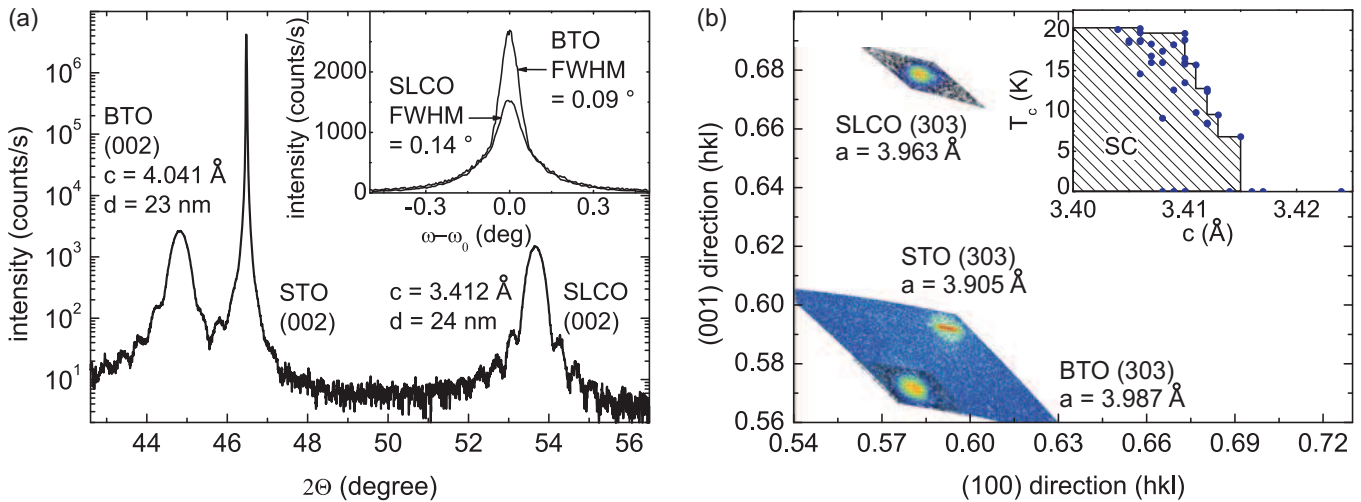}\fi
\caption{(Color online)
XRD data of SLCO on BTO/STO. (a) Main graph shows the (002) region of a $\Theta - 2 \Theta$ scan and inset shows the rocking curve of the SLCO (002) and BTO (002) peaks. (b) Main graph shows reciprocal space mapping of the (303) region. Inset displays $T_c$ vs $c$-axis parameter for a set of SLCO films grown under comparable conditions. Superconductivity (SC) is expected only to occur within the shaded region.
}
\label{fig:slcobtoxrd}
\end{figure*}
%
Figure \ref{fig:slcobtoxrd}~(a) shows an XRD $\Theta - 2 \Theta$ scan of a typical BTO-buffered SLCO film.
From the position of the SLCO (002) reflection, the $c$-axis parameter is calculated as 3.412\,\AA.
A set of comparable films with $T_c$ close to 20\,K (cf.~inset of Fig.~\ref{fig:slcobtoxrd}~(b)) revealed a mean value of $c_\mathrm{SLCO} = (3.408 \pm 0.002)\,$\AA, which is close to the value reported for Sr$_{0.9}$La$_{0.1}$CuO$_2$ films on DSO substrates ($\sim 3.410\,$\AA\,)\cite{Karimoto04}.
As deduced from Laue oscillations of the SLCO (002) peak and from narrow rocking curves (cf.~Fig.~\ref{fig:slcobtoxrd}~(a)), the films have a high crystalline quality and a flat surface.
Large angle $\Theta - 2 \Theta$ scans ($2\Theta = 0$\,-\,$90\,^\circ$) revealed only one phase, confirming that the SLCO film is single phase with IL crystal structure.
All films are aligned along the $a$-/$b$-axis of STO, as found by $\Phi$ scans around the sample normal (not shown here).
\\
With the aid of two-axes scans around the SLCO (303) reflection, as shown in the main graph of Fig.~\ref{fig:slcobtoxrd}~(b), we determined the in-plane lattice constant $a_\mathrm{SLCO} = 3.963\,$\AA.
The films with $T_c$ close to 20\,K exhibited a mean value of $a_\mathrm{SLCO} = (3.967 \pm 0.002)\,$\AA, which is somewhat larger than the reported value of 3.955\,\AA\, for Sr$_{0.1}$La$_{0.9}$CuO$_2$ films on DSO\cite{Karimoto04}.
However, as compared to the BTO buffer layers with a mean value of $a_\mathrm{BTO} = 3.994\,$\AA\, (cf.~sec.~\ref{subsubsec:btoxrd}), $a_\mathrm{SLCO}$ is rather small.
We explain this difference primarily by the interplay of lattice mismatch (between BTO and SLCO) inducing tensile strain, and excess oxygen inducing compressive strain.
Other influences, such as (off-)stoichiometry (different ionic radii) might also contribute to the final value of $a_\mathrm{SLCO}$.
Altogether, this results in SLCO films on BTO/STO with inferior tensile strain, as illustrated in the main graph of Fig.~\ref{fig:slcobtoxrd}~(b), where a shift of the SLCO (303) peak along the (100) direction with respect to the BTO (303) peak is visible.
\\
The inset of Fig.~\ref{fig:slcobtoxrd}~(b) shows the correlation of the transition temperature $T_c$ and the out-of-plane lattice constant $c_\mathrm{SLCO}$ for a set of SLCO films fabricated under comparable conditions.
No superconducting transition was observed for $c_\mathrm{SLCO} > 3.415\,$\AA\, and $T_c$ was highest for smallest $c_\mathrm{SLCO}$.
This observation is explained by the amount of incorporated excess oxygen, which expands the $c$-axis and hampers superconductivity, as described in sec.~\ref{sec:Introduction}.
We explain data points well inside the shaded region by non-ideal process conditions, e.g.~by too high vacuum annealing temperature, leading to oxygen vacancies in the CuO$_2$ planes, which again decreases $T_c$.
We note, that more films followed this trend, however, for clarity we only show data points for films prepared under comparable conditions as described in sec.~\ref{subsec:slcobtosto}.

\subsubsection{Rutherford backscattering spectroscopy}
\label{subsubsec:slcobtorbs}

\begin{figure}[tb]
\centering
\ifgraph\includegraphics[width=0.95\columnwidth]{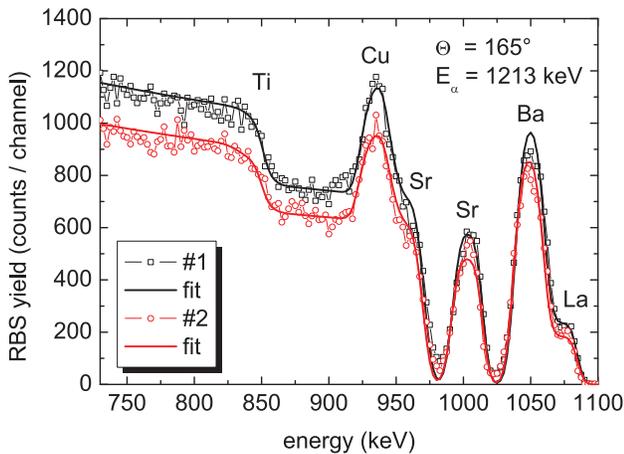}\fi
\caption{(Color online)
Two Rutherford backscattering spectra (open symbols) of an SLCO film on BTO/STO measured in two independent runs (\#1, \#2). Solid lines are numerical fits to the data. The determined stoichiometry is given in table \ref{tab:slcobtorbs}.
}
\label{fig:slcobtorbs}
\end{figure}
%
To determine the stoichiometry of the BTO-buffered SLCO films, we performed RBS on two samples\cite{Diebold10}.
The first sample had a thickness of 24\,nm and a $T_c$ of 14.9\,K.
As determined by XRD, the lattice parameters were $a_\mathrm{SLCO} = 3.963\,$\AA\, and $c_\mathrm{SLCO} = 3.411\,$\AA, i.e. the film had a slight amount of excess oxygen.
Simulations showed that the separation of the La and the Ba peak in the backscattering spectrum is stronger for lower $\alpha$-energies\cite{Diebold10,Simnra}.
For this reason, measurements were performed at 1213\,keV.
The resolution of the detector at this energy is 20\,keV/1213\,keV = 1.6\,\%.
Two independent measurements yielded consistent data and were performed to allow for a deduction of the double standard deviation 2$\sigma$.
The data was fitted numerically using a simplex algorithm\cite{Simnra}.
The measurements are shown in Fig.~\ref{fig:slcobtorbs} and the results are summarized in table \ref{tab:slcobtorbs}.
Note, that the Sr-peak (at $\sim 1000\,$keV) stemming from the SLCO film is separate from the Sr-box (ending at $\sim 970\,$keV) stemming from the STO substrate because of the energy loss of $\alpha$-particles crossing the intermediate BTO film.

\begin{table}
\caption{Areal density (in units of $10^{15}$ atoms/cm$^2$) as determined by numerical fits of RBS data measured on a BTO-buffered SLCO film on STO. The two independent measurements are labeled as \#1 and \#2. Errors of the numerical fits are denoted in parentheses. The error of the mean value is the double standard deviation $2\sigma$.}
\begin{tabular}{cccc}
 \hline
 \hline
\hspace*{0.5cm} meas. & \hspace*{0.5cm} Sr & \hspace*{0.5cm} La & \hspace*{0.5cm} Cu \\
\hline
\hspace*{0.3cm} \#1 		& \hspace*{0.3cm} $29.34(9.78)$ & \hspace*{0.3cm} $5.40(0.13)$ & \hspace*{0.3cm} $\,\,\,36.21(12.07)$ \\
\hspace*{0.3cm} \#2 		& \hspace*{0.3cm} $29.46(9.77)$ & \hspace*{0.3cm} $5.11(0.14)$ & \hspace*{0.3cm} $\,\,\,34.57(11.47)$ \\
\hspace*{0.3cm} mean 	& \hspace*{0.3cm} $29.39[0.11]$ & \hspace*{0.3cm} $5.25[0.29]$ & \hspace*{0.3cm} $35.39[1.64] $ \\
\hline
\hline
\end{tabular}
\label{tab:slcobtorbs}
\end{table}
%
For stoichiometric SLCO with a sum formula Sr$_{1-x}$La$_x$CuO$_2$, the sum of Sr and La atoms equals the number of Cu atoms.
In our case this is $(29.4 + 5.3)\times 10^{15}\,$atoms/cm$^2 = 34.7\times 10^{15}\,$atoms/cm$^2$, which is very close to the determined amount of Copper ($35.4\times 10^{15}$\,atoms/cm$^2$).
Therefore, an off-stoichiometry between (Sr/La) and Cu can be excluded within experimental accuracy.
Thus, from RBS data we extract a sum formula Sr$_{0.84}$La$_{0.16}$CuO$_2$, showing that the sample is overdoped with respect to the doping level of the target ($x = 0.125$).
The second sample examined by RBS yielded similar results with $x = 0.14$.
Note, that RBS fits have only been performed on the high energy part of the spectrum where the peaks of Ti, Cu, Sr, Ba and La are visible and not for the low energy part where the oxygen peak is visible.
To summarize our RBS measurements, we determined the doping level of BTO-buffered SLCO films as $x \approx 0.15$, i.e.~the samples are overdoped with respect to the doping level reported for SLCO films exhibiting maximum $T_c$ ($x_\mathrm{opt} = 0.10$)\cite{Karimoto01} and to the solid solution level reported for polycrystalline bulk SLCO ($x_\mathrm{sol} = 0.10$)\cite{Er92}.
Note, that the maximum $T_c$ of our SLCO films was 22\,K, which can be explained by overdoping.
However, other influences such as excess oxygen or defects may also contribute to a reduction of $T_c$ from its maximum value of 43\,K.

\subsubsection{Electric transport measurements}
\label{subsubsec:slcobtoelec}

\begin{figure}[tb]
\centering
\ifgraph\includegraphics[width=1.00\columnwidth]{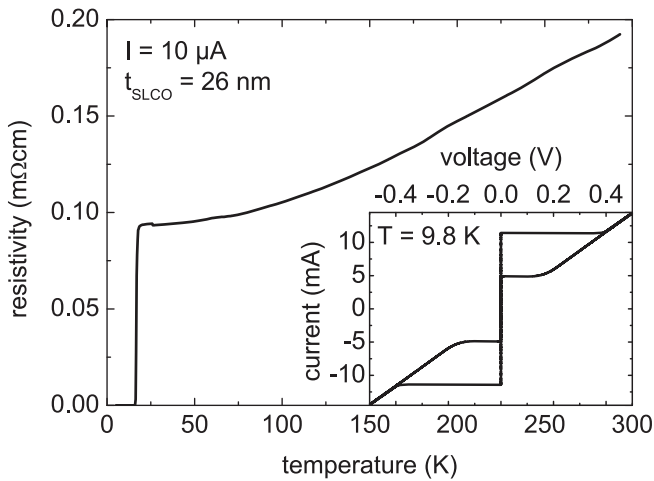}\fi
\caption{
Resistivity vs temperature of an unpatterned SLCO film on BTO/STO. Inset shows the current-voltage characteristics of a $40\,\mu$m wide bridge at $T = 9.8\,$K.
}
\label{fig:slcobtort}
\end{figure}
%
Figure \ref{fig:slcobtort} shows a typical $\rho(T)$ curve of an unpatterned SLCO film on BTO/STO.
The superconducting transition at $T_c = 16.8\,$K has a width of $\Delta T_c \approx 1\,$K.
The highest $T_c$ achieved was 22\,K.
The room temperature resistivity is $\rho_\mathrm{300K} \approx 0.2\,\mathrm{m}\Omega$cm, which is comparable to the values reported for YBa$_2$Cu$_3$O$_{7-\delta}$ single crystals and high quality thin films.
We further observed an almost linear $\rho(T)$ dependence for $T \,\gapprox\, 100\,$K, which was explained by scattering due to spin fluctuations in the CuO$_2$ planes in case of YBa$_2$Cu$_3$O$_{7-\delta}$\cite{Ito93}.
The residual resistance ratio is $\mathrm{RRR} \approx 2$, which is somewhat lower than the value reported for optimally doped Sr$_{0.90}$La$_{0.10}$CuO$_2$ films on KTO, where $\mathrm{RRR} \approx 3$\cite{Karimoto01}.
Possible explanations for reduced $T_c$ and RRR are excess oxygen in the (Sr/La) planes, oxygen deficiency in the CuO$_2$ planes, lattice defects, impurities, or off-stoichiometry.
As shown in sec.~\ref{sec:discussion}, both excess oxygen and oxygen deficiency lead to semiconducting or insulating $\rho(T)$ behavior.
Moreover, lattice defects or impurities lead to a finite resistance at low temperatures, where electron-phonon scattering is negligible, leading to a reduced RRR.
After $\rho(T)$ measurement, the SLCO film was patterned to allow for determination of the critical current density $j_c(T)$.
The $\rho(T)$ behavior measured with these bridges (not shown here) coincided with the $\rho(T)$ behavior determined by van der Pauw measurement prior to patterning, which confirms that patterning did not affect the properties of SLCO.
We measured $I(V)$ curves in the temperature range $4.2\,\mathrm{K} \le T \le T_c$, as shown in the inset of Fig.~\ref{fig:slcobtort} for $T = 9.8\,$K.
The $I(V)$ curves were hysteretic, probably due to heating.
At $T = 4.2\,$K, we found a critical current density $j_c(4.2\,K) = 2.1 \times 10^{6}\,$A/cm$^2$, which is 1 to 2 orders of magnitude lower than the values reported for YBCO at $T = 4.2\,$K but coincides with that of $T'$-compounds such as Nd$_{2-x}$Ce$_x$CuO$_4$\cite{Nishizaki94}.

\subsection{Characterization of SLCO on KTO}
\label{subsec:charslcokto}

\subsubsection{RHEED \& atomic force microscopy}

\begin{figure*}[tb]
\centering
\ifgraph\includegraphics[width=0.95\textwidth]{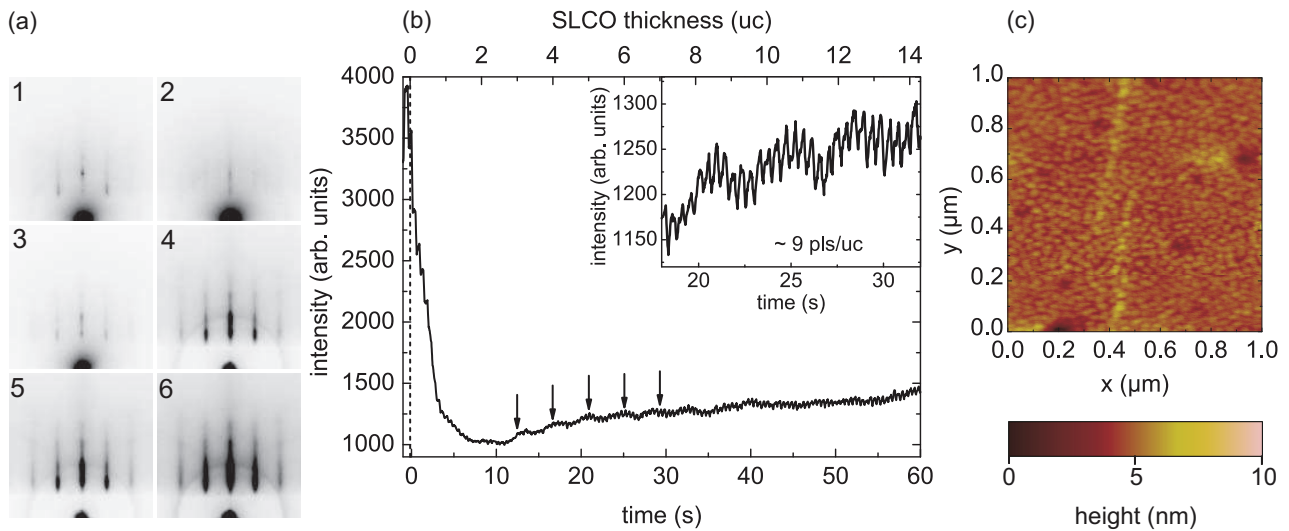}\fi
\caption{(Color online)
(a) RHEED patterns during preparation of SLCO on KTO. 1: KTO substrate before deposition, 2: SLCO after growth of $\sim 2$ unit cells, 3: end of SLCO deposition, 4: begin of SLCO vacuum annealing, 5: end of SLCO vacuum annealing, 6: sample at room temperature. (b) Intensity oscillations of the RHEED specular spot during growth of SLCO on KTO (marked by arrows). Deposition started at $t = 0$. Inset presents a zoom, showing growth of $\sim 3\,$uc in detail. (c) AFM image of the SLCO surface. The root mean square roughness is 0.35\,nm and the maximum step height is 1.0\,nm, corresponding to 3\,uc SLCO. The darker regions stem from holes in the as-received KTO substrate.
}
\label{fig:slcoktorheed}
\end{figure*}
%
During initial growth of SLCO on KTO, typically $(10 \pm 5)$ RHEED oscillations could be observed, as displayed in Fig.~\ref{fig:slcoktorheed}~(b).
This value is somewhat lower than what is found for SLCO on BTO/STO (cf.~sec.~\ref{subsubsec:slcobtorheedafm}).
A possible explanation is given by the fact that KTO was not vacuum annealed before deposition of SLCO, leading to enhanced island growth (increased step density) due to worse substrate-film interface and thus to a faster disappearance of RHEED oscillations.
The oscillations revealed a growth rate of $(8.5 \pm 1.0)$\,pls/uc, corresponding to a film thickness of $t_\mathrm{SLCO} = (34 \pm 4)\,$nm for 850 pulses.
At the end of deposition, the RHEED pattern revealed a 2D surface with fade 3D admixtures, possibly due to 3D islands or small droplets on the surface (cf.~Fig.~\ref{fig:slcoktorheed}~(a)).
Together with the intensity evolution of the specular spot, we can thus identify a Stranski-Krastanov growth mode\cite{Stranski39}.
Note, that for SLCO films on KTO, no intermediate streaks were observed, which is different from what we found for SLCO films on BTO/STO.
\\
Figure \ref{fig:slcoktorheed}~(c) shows an AFM image of SLCO on KTO.
The root mean sqaure roughness is $RMS = 0.35\,$nm and the maximum step height is $\sim 1.0\,$nm, i.e.~the maximum roughness is caused by asperities of 3\,uc.
The morphology of SLCO on KTO is thus comparable to the morphology of SLCO on BTO/STO.
Note, that extra structures visible in the AFM image, i.e.~ bright lines and dark holes, stem from the KTO substrate, which was not annealed prior to deposition.

\subsubsection{X-ray diffraction}

\begin{figure*}[tb]
\centering
\ifgraph\includegraphics[width=0.95\textwidth]{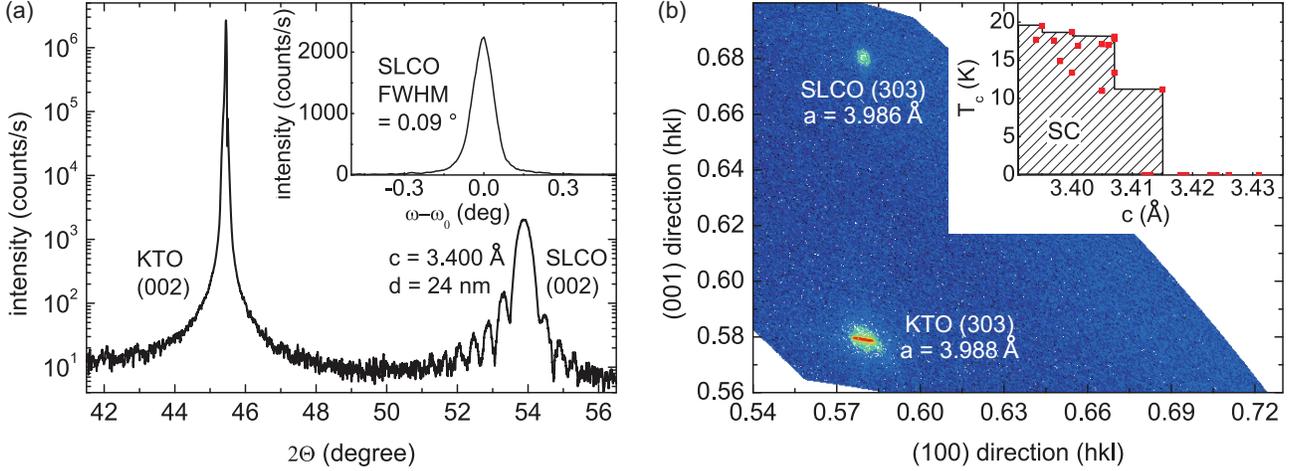}\fi
\caption{(Color online)
XRD data of SLCO on KTO. (a) Main graph shows the (002) region of a $\Theta - 2 \Theta$ scan and inset shows the rocking curve of the SLCO (002) peak. (b) Main graph shows reciprocal space mapping of the (303) region. Inset displays $T_c$ vs $c$-axis parameter for a set of SLCO films grown under comparable conditions. Superconductivity (SC) is only observed within the shaded region.
}
\label{fig:slcoktoxrd}
\end{figure*}
%
A typical XRD $\Theta - 2 \Theta$ scan of SLCO on KTO is shown in the main graph of Fig.~\ref{fig:slcoktoxrd}~(a), indicating that the film is single phase with an IL crystal structure.
About 10 Laue oscillations can be identified, pointing to a uniform crystal structure along the $c$-axis and a flat film surface.
Moreover, the narrow rocking curve with $\mathrm{FWHM} = 0.09\,^\circ$ demonstrates high crystalline quality of the film.
The $c$-axis parameter is calculated as 3.400\,\AA.
For comparable films, exhibiting $T_c$ close to 20\,K, the mean value was $c_\mathrm{SLCO} = (3.401 \pm 0.005)\,$\AA, which coincides well with the value $c = (3.400 \pm 0.003)\,$\AA\, reported for Sr$_{0.9}$La$_{0.1}$CuO$_2$ films on KTO\cite{Karimoto02,Karimoto04}.
\\
Reciprocal space mapping around the (303) reflection was used to determine the in-plane lattice parameter $a_\mathrm{SLCO}$.
As displayed in Fig.~\ref{fig:slcoktoxrd}~(b), the SLCO film is under tensile epitaxial strain with $a_\mathrm{SLCO} = 3.986\,$\AA, which is close to the substrate parameter $a_\mathrm{KTO} = 3.988\,$\AA.
The mean value for films with $T_c$ close to 20\,K was $a_\mathrm{SLCO} = (3.982 \pm 0.004)\,$\AA, which is somewhat larger than the value $(3.972 \pm 0.006)\,$\AA\, reported for Sr$_{0.9}$La$_{0.1}$CuO$_2$ films on KTO\cite{Karimoto02,Karimoto04}.
\\
The inset of Fig.~\ref{fig:slcoktoxrd}~(b) shows the dependence of $T_c$ on the lattice constant $c_\mathrm{SLCO}$. As found for SLCO on BTO/STO (cf.~sec.~\ref{subsubsec:slcobtoxrd}), an upper limit for $c_\mathrm{SLCO}$ where superconductivity occurs can be identified empirically ($c_\mathrm{SLCO} \leq 3.415\,$\AA).

\begin{figure}[tb]
\centering
\ifgraph\includegraphics[width=1.00\columnwidth]{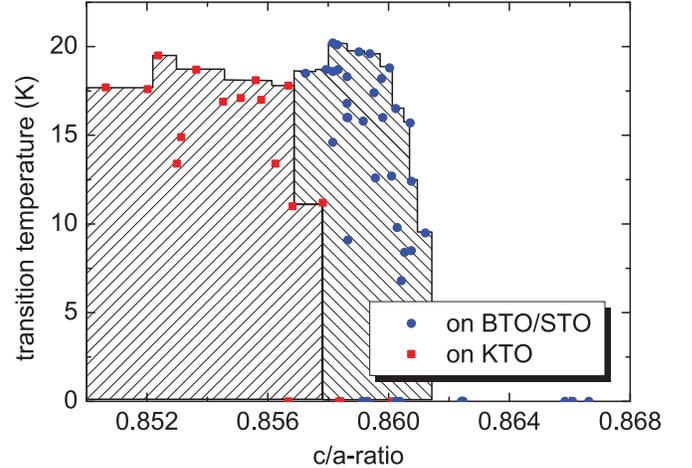}\fi
\caption{(Color online)
Transition temperature $T_c$ vs $c/a$-ratio for SLCO on BTO/STO and on KTO. The shaded regions, where superconductivity is found to occur, clearly reveal two different regimes of tensile strain.
}
\label{fig:tcca}
\end{figure}

\begin{table}
\caption{Comparison of typical XRD data for SLCO films grown on different substrates. All data refer to the (002) reflection. The FWHM corresponds to the rocking curve and the number of Laue oscillations (osc.) to the $\Theta - 2 \Theta$ scan. $c_\mathrm{SLCO}$ and $a_\mathrm{SLCO}$ are the mean values of films with $T_c \approx 20\,$K.}
\begin{tabular}{lccccc}
 \hline
 \hline
substrate	& \hspace{0.2cm} intensity 		& \hspace{0.2cm} FWHM 			& \hspace{0.2cm} osc.	& \hspace{0.2cm} $c_\mathrm{SLCO}$		& \hspace{0.2cm} $a_\mathrm{SLCO}$ 	\\
 					& \hspace{0.2cm}(cps) 				& \hspace{0.2cm} (deg) 			& \hspace{0.2cm}(\#)					& \hspace{0.2cm} (\AA)	& \hspace{0.2cm} (\AA)	\\
\hline
BTO/STO 	& \hspace{0.2cm} $\sim 1000$ & \hspace{0.2cm} $\sim 0.15$ & \hspace{0.2cm} $\sim 4$			& \hspace{0.2cm} 3.408	& \hspace{0.2cm} 3.967 	\\
KTO 			& \hspace{0.2cm} $\sim 2000$ & \hspace{0.2cm} $\sim 0.11$ & \hspace{0.2cm} $\sim 8$			& \hspace{0.2cm} 3.401	& \hspace{0.2cm} 3.982 	\\
\hline
\hline
\end{tabular}
\label{tab:slcocompare}
\end{table}
%
We now compare the XRD data of SLCO films on BTO/STO (cf.~sec.~\ref{subsubsec:slcobtoxrd}) with XRD data of SLCO films on KTO.
Table \ref{tab:slcocompare} comprises a summary of XRD data typically measured for SLCO films grown on both types of substrates.
It is obvious, that SLCO films on KTO show higher peak intensities, narrower rocking curves and more Laue oscillations, i.e.~better crystalline quality.
More important, the $a$-axis ($c$-axis) is larger (smaller) for SLCO films on KTO, resulting in a smaller ratio $c/a = 0.854$ for SLCO films on KTO, compared to 0.859 for SLCO films on BTO/STO, i.e.~SLCO films on KTO are more tensile strained.
It is thus remarkable that both kinds of samples exhibit similar electric transport properties (cf.~sec.~\ref{subsubsec:transportslcokto}), in particular the same maximum $T_c \approx 20\,$K, whereupon tensile strain was believed to enhance electron doping of the CuO$_2$-planes and therefore increase $T_c$\cite{Er91,Sugii93,Karimoto01,Markert03}.
For illustration, Fig.~\ref{fig:tcca} shows the correlation of the transition temperature $T_c$ vs $c/a$-ratio for SLCO films deposited on both kinds of substrates.
It is obvious, that SLCO films on BTO/STO are superconducting for $c/a \,\lapprox\, 0.862$, whereas SLCO films on KTO are superconducting for $c/a \,\lapprox\, 0.858$.
To explain this difference, we refer to the work of Karimoto \textit{et al.}:
On the one hand, they showed that \textit{compressively strained} SLCO films on STO are not superconducting\cite{Sugii93,Karimoto01}, whereas \textit{tensile strained} SLCO films on KTO are superconducting with $T_c^\mathrm{zero} = 39\,$K.
On the other hand, they found that \textit{relaxed} SLCO films on DSO exhibit slightly improved electric transport properties with $T_c^\mathrm{zero} = 41\,$K\cite{Karimoto04}.
Thus, whereas compressive strain hampers superconductivity, both relaxed and tensile strained films exhibit comparable superconducting properties.
Therefore, despite exhibiting different degrees of tensile strain, it is in accordance with literature that our SLCO films show comparable electric transport properties.

\subsubsection{Electric transport measurements}
\label{subsubsec:transportslcokto}

\begin{figure}[tb]
\centering
\ifgraph\includegraphics[width=1.00\columnwidth]{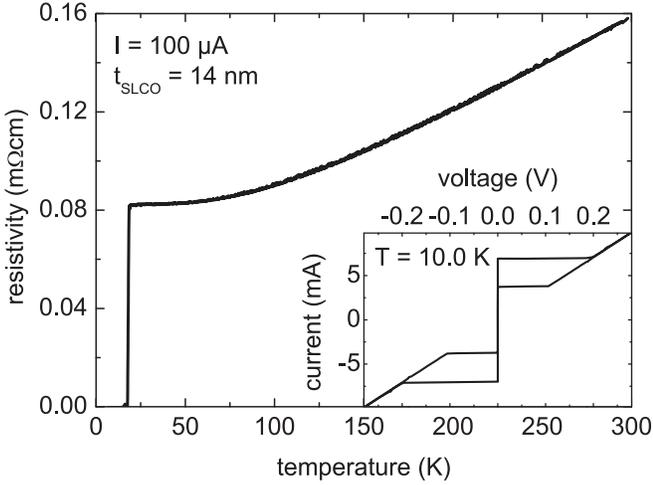}\fi
\caption{
Resistivity vs temperature of an unpatterned SLCO film on KTO. Inset shows current-voltage characteristics of a 40\,$\mu$m wide bridge at $T = 10.0\,$K.
}
\label{fig:slcoktort}
\end{figure}
%
Figure \ref{fig:slcoktort} shows $\rho(T)$ of an SLCO film on KTO with $T_c = 18.0\,$K.
The transition width is $\Delta T_c \approx 1\,$K.
The room temperature resistivity is $\rho_\mathrm{300K} = 0.16\,$m$\Omega$cm and $\mathrm{RRR} \approx 2$.
We observed hysteretic $I(V)$ curves, as shown in the inset of Fig.~\ref{fig:slcoktort} for $T = 10\,$K.
At $T = 4.2\,$K, we found a critical current density $j_c(4.2\,K) = 2.0 \times 10^6\,$A/cm$^2$, which is comparable to the value found for SLCO on BTO/STO, cf.~sec.~\ref{subsubsec:slcobtoelec}.
Altogether, despite of slightly superior crystalline properties of SLCO films on KTO, their electric transport properties are quite similar to those of SLCO films on BTO/STO.

\section{Discussion of process parameters and vacuum annealing}
\label{sec:discussion}

In this chapter, the influence of different process parameters on structural and electric properties of BTO and SLCO films are discussed.

\subsection{Influence of process parameters on BTO}
\label{subsec:procbto}

\subsubsection{Substrate temperature}
\label{subsubsec:btosubsttemp}

We examined the influence of the substrate temperature $T_\mathrm{BTO}$ during deposition of BTO on its final lattice parameters.
BTO films grown at comparable temperatures $T_\mathrm{BTO} = (670 \pm 15)\,^\circ$C and oxygen pressure $p_\mathrm{O_2} = 10\,$Pa were analyzed by means of XRD.
The lattice parameters were determined as $a_\mathrm{BTO} = (3.994 \pm 0.003)\,$\AA\, and $c_\mathrm{BTO} = (4.021 \pm 0.003)\,$\AA.
Another set of samples grown at higher substrate temperature $T_\mathrm{BTO} = (760 \pm 15)\,^\circ$C revealed lattice parameters of $a_\mathrm{BTO} = (3.988 \pm 0.006)\,$\AA\, and $c_\mathrm{BTO} = (4.033 \pm 0.007)\,$\AA.
Thus, for increasing substrate temperature, the $c$-axis increases and the $a$-axis decreases.
As known from literature, reduction of BTO results in an increased $c/a$-ratio.
We therefore interpret our result as follows.
At high $T$ oxygen is more mobile than at low $T$, leading to an enhanced desorption of oxygen already during film growth.
The films deposited at high $T$ are consequently more reduced and have a larger $c/a$-ratio than those grown at low $T$.
\\
Moreover, by means of RHEED, we found that a superstructure emerged when BTO was deposited at $T_\mathrm{BTO} \,\lapprox\, 650\,^\circ$C, which may be the result of surface defects.
Therefore, for the deposition pressure $p_\mathrm{O_2} = 10\,$Pa, $T_\mathrm{BTO}$ values higher than 650$\,^\circ$C were used.

\subsubsection{Vacuum annealing}

As described in sec.~\ref{subsec:btosto}, BTO films were annealed at $T_a \approx 900\,^\circ$C in oxygen ($p_\mathrm{O_2} = 10\,$Pa) for $t_a = 15\,$min and subsequently at the same temperature in vacuum ($p_\mathrm{vac} \,\lapprox\, 10^{-5}\,$Pa) for $t_a = 30\,$min.
To examine the influence of this vacuum annealing step on the lattice constants, we also fabricated some reference samples which were only annealed in oxygen but not in vacuum.
Those films revealed a larger in-plane lattice constant $a_\mathrm{BTO} = (3.992 \pm 0.005)\,$\AA\, and a smaller out-of-plane lattice constant $c_\mathrm{BTO} = (4.026 \pm 0.013)\,$\AA\, as compared to the films \textit{with} additional vacuum annealing ($a_\mathrm{BTO} = (3.988 \pm 0.006)\,$\AA\, and $c_\mathrm{BTO} = (4.033 \pm 0.007)\,$\AA).
From this observation we can conclude that reduction by vacuum annealing leads to an increase of the $c/a$-ratio, in accordance with literature.
Furthermore, it supports the interpretation given in sec.~\ref{subsubsec:btosubsttemp}.

\subsection{Influence of process parameters on SLCO}
\label{subsec:procslco}

\subsubsection{Excimer laser energy}
\label{subsubsec:laserenergy}

\begin{figure}[tb]
\centering
\ifgraph\includegraphics[width=0.95\columnwidth]{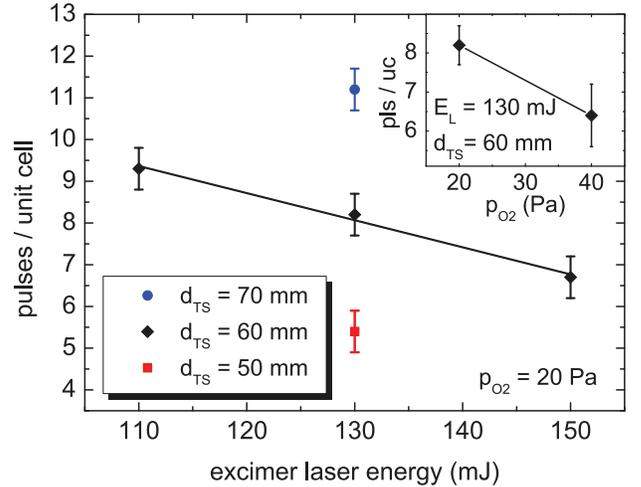}\fi
\caption{(Color online)
Number of laser pulses needed to deposit one monolayer (determined by RHEED oscillations) of SLCO on BTO/STO at an oxygen pressure $p_\mathrm{O_2} = 20\,$Pa vs excimer laser energy. The solid line is a linear fit to the data acquired at a target-to-substrate distance $d_\mathrm{TS} = 60\,$mm. Inset shows the dependence of the growth rate on the oxygen pressure (for fixed $E_L = 130\,$mJ and $d_\mathrm{TS} = 60\,$mm).
}
\label{fig:slcobtoplsuc}
\end{figure}
%
In this section, we report on the influence of the excimer laser energy $E_L$ on SLCO films.
For that purpose, we fixed all other parameters and varied $E_L$.
By analyzing the RHEED oscillations during deposition of $\sim 30$ SLCO films on BTO/STO, we found that the number of pulses needed to deposit one unit cell SLCO (pls/uc) decreased approximately linearly with increasing laser energy.
For a target-to-substrate distance $d_\mathrm{TS} = 60\,$mm, the values determined during deposition were ($9.3 \pm 0.5)\,$pls/uc at 110\,mJ, ($8.2 \pm 0.5)\,$pls/uc at 130\,mJ, and ($6.7 \pm 0.5)\,$pls/uc at 150\,mJ, as displayed in Fig.~\ref{fig:slcobtoplsuc}.
It is obvious that higher $E_L$ yields a higher plasma density and thus an increased growth rate.
\\
To analyze the impact of varying laser energy on the lattice parameters, we compare two representative SLCO films on BTO/STO, prepared identically except for $E_L$.
One film was prepared with $E_L = 130\,$mJ and the other with $E_L = 150\,$mJ.
As the growth rate increased with increasing $E_L$, 550 and 400 pulses were deposited, respectively to end up with the same film thickness.
The resulting lattice parameters were $c_\mathrm{SLCO} = 3.410\,$\AA\, and $a_\mathrm{SLCO} = 3.965\,$\AA\, for $E_L = 130\,$mJ as well as $c_\mathrm{SLCO} = 3.419\,$\AA\, and $a_\mathrm{SLCO} = 3.962\,$\AA\, for $E_L = 150\,$mJ, i.e.~the $c/a$-ratio increased with increasing laser energy.
This effect can be attributed to an enhanced incorporation of excess oxygen:
A laser pulse with higher $E_L$ creates more high energetic particles in the plasma.
Therefore, more oxygen is activated in the surrounding process gas and incorporated into the SLCO, which would well explain the increased $c/a$-ratio.
However, we can only speculate about this dependence and further examination is needed to verify this assumption.

\subsubsection{Target-to-substrate distance}
\label{subsubsec:dts}

Another degree of freedom is given by the distance $d_\mathrm{TS}$ between target and substrate.
It is obvious that an increase of $d_\mathrm{TS}$ should reduce the growth rate because of the dilution of the plasma in outer regions.
Indeed, we observed such behavior by analyzing RHEED oscillations of 25 SLCO films deposited on BTO/STO at similar conditions but with varying $d_\mathrm{TS}$.
The values for $d_\mathrm{TS} = 50\,$mm, 60\,mm, and 70\,mm at fixed $E_L = 130\,$mJ were $(5.4 \pm 0.5)\,$pls/uc, $(8.2 \pm 0.5)\,$pls/uc, and $(11.2 \pm 0.5)\,$pls/uc, respectively, as displayed in Fig.~\ref{fig:slcobtoplsuc}.
As discussed in sec.~\ref{subsubsec:laserenergy}, higher $E_L$ probably leads to an enhanced incorporation of oxygen.
We want to interpret this result in another way:
higher $E_L$ leads to an expansion of the plume and consequently to a change of the \textit{relative} position of the substrate within the plume.
Actually, this corresponds to a situation, where the laser energy was kept constant but the substrate was moved closer to the target.
Thus, we expect that the oxygen concentration should increase with decreasing $d_\mathrm{TS}$, too.
To check this idea, we deposited a film at $d_\mathrm{TS} = 50\,$mm and another at $d_\mathrm{TS} = 70\,$mm and compared their lattice parameters.
To end up with the same film thickness ($t_\mathrm{SLCO} \approx 23\,$nm), 350 and 770 pls were ablated, respectively.
The resulting lattice parameters were $c_\mathrm{SLCO} = 3.410\,$\AA\, and $a_\mathrm{SLCO} = 3.966\,$\AA\, as well as $c_\mathrm{SLCO} = 3.406\,$\AA\, and $a_\mathrm{SLCO} = 3.969\,$\AA, supporting the abovementioned idea.
To conclude, we found that smaller $d_\mathrm{TS}$ results in an increased $c/a$-ratio, probably due to enhanced incorporation of excess oxygen.
\\
To remove excess oxygen, we prepared a sample identical to that deposited at $d_\mathrm{TS} = 50\,$mm but with additional in-situ vacuum annealing ($t_a = 40\,$min) at lower temperature ($T_a \approx 350\,^\circ$C).
The lattice constants of this sample were $c_\mathrm{SLCO} = 3.408\,$\AA\, and $a_\mathrm{SLCO} = 3.969\,$\AA, which is close to the parameters of the sample prepared at 70\,mm \textit{without} additional low temperature annealing.
This confirms that excess oxygen was indeed responsible for the increase of the $c/a$-ratio, which is both dependent on $E_L$ and on $d_\mathrm{TS}$.
Finally, we want to mention that the Sr and/or La concentration, i.e.~the doping level $x$, is probably also dependent on $E_L$ and $d_\mathrm{TS}$:
In a naive approach, the heavier La ($m_\mathrm{La} = 138.91\,$u)\cite{CRC71} should dominate the inner regions of the plasma because of its high inertia, whereas the lighter Sr ($m_\mathrm{Sr} = 87.62\,$u)\cite{CRC71} should dominate the outer regions.
Yet, a validation of this idea goes beyond the scope of this work.

\subsubsection{Deposition pressure}

To analyze the influence of oxygen pressure on the SLCO films, we prepared films at $p_\mathrm{O_2} = 20\,$Pa and $p_\mathrm{O_2} = 40\,$Pa and compared their properties.
RHEED oscillations of 25 SLCO films on BTO/STO revealed an increase of the growth rate with increasing $p_\mathrm{O_2}$.
The values were $(8.2 \pm 0.5)\,$pls/uc and $(6.4 \pm 0.5)\,$pls/uc, respectively, which is illustrated in the inset of Fig.~\ref{fig:slcobtoplsuc}.
However, not only the growth rate but also the lattice constants changed.
The lattice parameters of a film grown at $p_\mathrm{O_2} = 20\,$Pa were $c_\mathrm{SLCO} = 3.408\,$\AA\, and $a_\mathrm{SLCO} = 3.969\,$\AA, while for a film grown at $p_\mathrm{O_2} = 40\,$Pa these values were $c_\mathrm{SLCO} = 3.412\,$\AA\, and $a_\mathrm{SLCO} = 3.965\,$\AA.
Again, we attribute this increase of the $c/a$-ratio to enhanced incorporation of excess oxygen, which is plausible, because higher oxygen pressure provides more active oxygen during film growth.

\subsection{Vacuum annealing}
\label{subsec:vacann}

\subsubsection{Vacuum annealing time}
\label{subsubsec:annealtime}

\begin{figure*}[tb]
\centering
\ifgraph\includegraphics[width=1.00\textwidth]{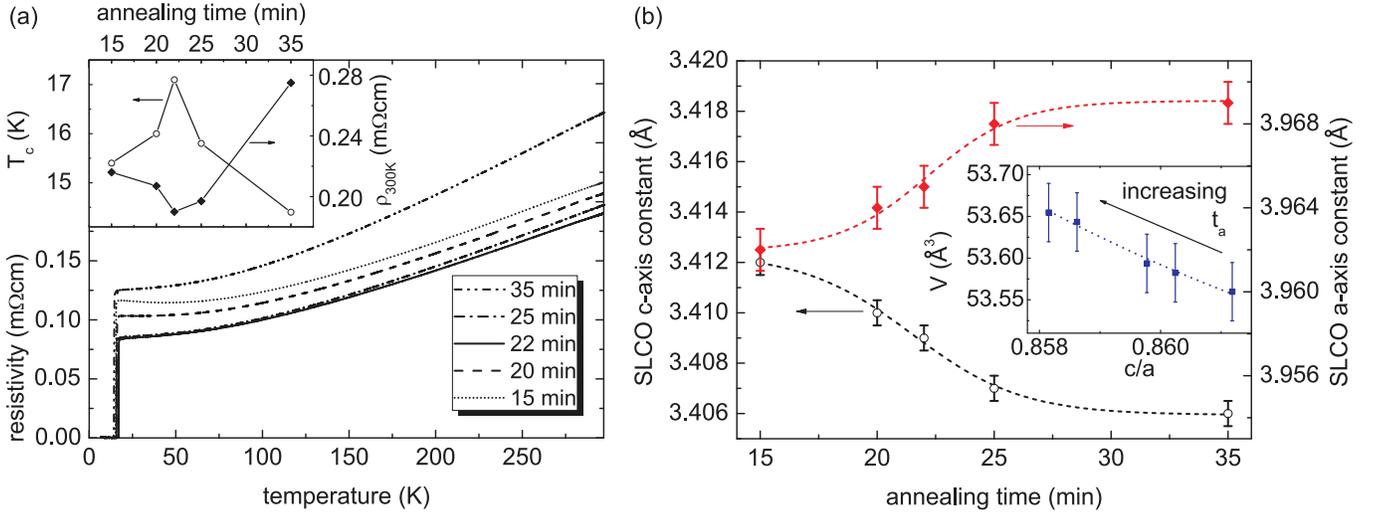}\fi
\caption{(Color online)
(a) Resistivity vs temperature of SLCO films deposited on BTO/STO with varying vacuum annealing time $t_a$. Inset shows the dependence of the transition temperature and of the room temperature resistivity vs $t_a$. (b) Dependence of the lattice parameters $a_\mathrm{SLCO}$ and $c_\mathrm{SLCO}$ vs $t_a$. Inset shows the the evolution of the unit cell volume $V = a_\mathrm{SLCO}^2 \times c_\mathrm{SLCO}$ with $t_a$. Lines are guides to the eyes.
}
\label{fig:slcobtoannt}
\end{figure*}
%
A series of SLCO films deposited on BTO/STO under similar conditions but with varying vacuum annealing time $t_a$ showed that there is an optimum annealing time $t_a^\mathrm{opt} = 22\,$min.
As shown in Fig.~\ref{fig:slcobtoannt}~(a), $T_c$ increases with increasing annealing time and reaches a maximum value of $17.1\,$K after 22\,min, before decreasing again.
The room temperature resistivity $\rho_\mathrm{300K}$ shows the opposite behavior with a minimum at $t_a^\mathrm{opt} = 22\,$min, which we explain as follows:
For $t_a < t_a^\mathrm{opt}$, excess oxygen is removed from SLCO.
As excess oxygen forms O$^{2-}$ ions on interstitial sites, it traps free charge carriers from the CuO$_2$ planes, which increases the resistivity and suppresses superconductivity.
Therefore, $\rho_\mathrm{300K}$ decreases and $T_c$ increases with the removal of excess oxygen, as reported before\cite{Karimoto02,Li09}.
We further observed the reduction process by analyzing the evolution of the lattice constants with proceeding $t_a$, as displayed in Fig.~\ref{fig:slcobtoannt}~(b).
In accordance with literature\cite{Karimoto02,Li09} we found that the $c$-axis decreases and the $a$-axis increases monotonically during reduction.
Furthermore, we found that the unit cell volume $V$ increases as well (cf.~inset of Fig.~\ref{fig:slcobtoannt}~(b)), which is a well-known behavior of various oxides when being reduced.\cite{Werner09}
The decrease of $T_c$ and the increase of $\rho_\mathrm{300K}$ for $t_a > 22\,$min is attributed to the formation of oxygen vacancies in the CuO$_{2}$-planes.
In hole-doped cuprates, oxygen vacancies in the CuO$_{2}$ planes are known to decrease the charge carrier concentration and to weaken the antiferromagnetic spin fluctuation/correlation of the $d$-electrons, suppressing $T_c$\cite{Gaojie01,Matsunaka05}.
For the IL compounds however, it has been shown that oxygen vacancies in the CuO$_{2}$ planes can even induce superconductivity\cite{Nie03,Nie03a}.
The idea is that those vacancies lead to electron doping of the CuO$_{2}$ planes, even without trivalent cation doping.
However, as oxygen vacancies are lattice defects at the same time, too strong reduction of the CuO$_{2}$ planes finally leads to suppression of superconductivity.

\begin{figure*}[tb]
\centering
\ifgraph\includegraphics[width=1.00\textwidth]{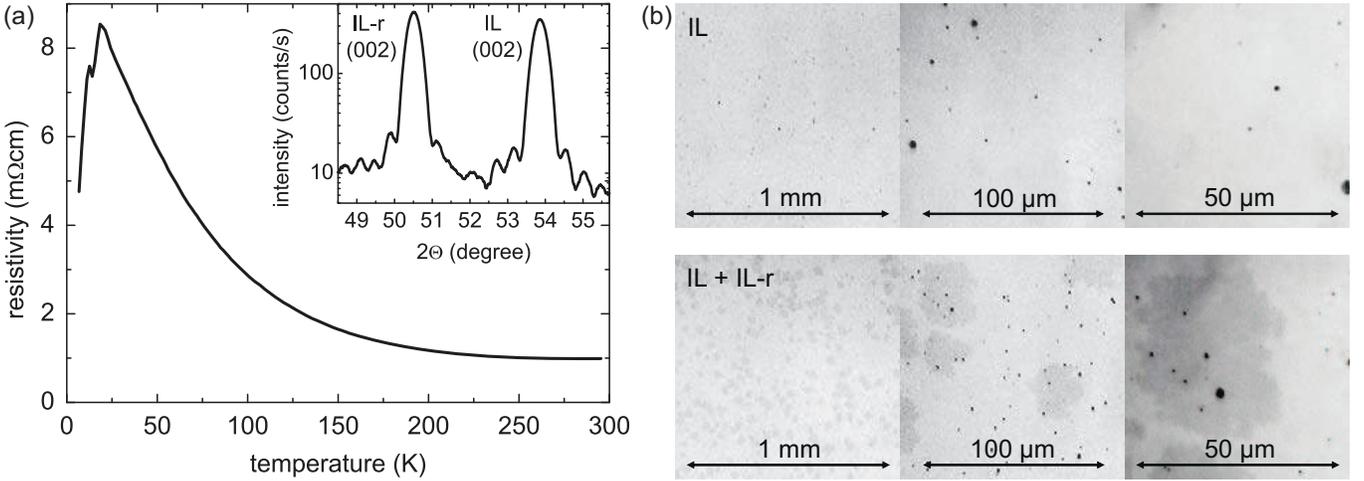}\fi
\caption{(Color online)
(a) Resistivity vs temperature of SLCO on BTO/STO, exhibiting both IL and IL-r phase, as confirmed by XRD $\Theta - 2 \Theta$ scans of the (002) region (cf.~inset). (b) Optical micrographs of a single phase IL film (top) and of a film with both phases, IL and IL-r (bottom). The regions with IL-r phase are visible as gray islands (cf.~text), droplets are visible as black spots.
}
\label{fig:ilr}
\end{figure*}
%
In the case of IL, a secondary phase can be formed if oxygen vacancies arrange in an ordered structure, which is called "infinite-layer-related" (IL-r) phase. 
It can be easily identified by means of XRD due to its elongated $c$-axis of $c_\mathrm{IL-r} \approx 3.6\,$\AA.
Figure \ref{fig:ilr}~(a) shows $\rho(T)$ and an XRD $\Theta - 2 \Theta$ scan of the (002) reflection of an SLCO film containing both phases, i.e.~the IL and IL-r phase.
The $c$-axis constants are determined as $c_\mathrm{SLCO} = 3.405\,$\AA\, and $c_\mathrm{IL-r} = 3.614\,$\AA.
The widths of the corresponding rocking curves are $\Delta \omega_\mathrm{SLCO} = 0.11\,^\circ$ and $\Delta \omega_\mathrm{IL-r} = 0.12\,^\circ$ (not shown here).
Laue oscillations of the (002) reflection are visible for both phases, allowing to determine their thickness.
For each phase we found a thickness of $(22 \pm 1)\,$nm, which is close to the total film thickness determined by RHEED oscillations ($t_\mathrm{SLCO} \approx 25\,$nm).
This implies that both phases coexist side by side and not on top of each other.
We reported on the same observation in a recent paper\cite{Tomaschko11}.
In conclusion, the formation of the IL-r phase is a process that develops laterally on distinct nuclei and not along the film normal.
Furthermore, we found that the IL and the IL-r phase can be imaged by optical microscopy due to different optical reflection.
Figure \ref{fig:ilr}~(b) shows optical micrographs of a single phase SLCO film (with IL crystal structure, top images) and of a double phase SLCO film (with IL and IL-r crystal structure, bottom images).
Whereas on top of the single phase film only droplets are visible (black spots with a diameter of a few $\mu$m), the double phase film shows several gray regions with an average size of $\sim 30 \times 30\,\mu$m$^2$.
As these regions were always correlated to signatures of the IL-r phase in XRD scans, we identify them as the IL-r phase.
Furthermore, the micrographs verify the proposed lateral formation process of the IL-r phase, as deduced from XRD measurements.
As illustrated in Fig.~\ref{fig:ilr}~(a), the IL-r phase is semiconducting, which supports the findings of Zhou \textit{et al.}\cite{Zhou93} but contradicts the interpretation of Karimoto \textit{et al.}\cite{Karimoto04,Karimoto04a} who suggest metallic behavior for the IL-r phase.
As in our sample the IL-r and the IL phase coexist side by side, we can also observe the fingerprint of the latter as onset of superconductivity at $T \approx 18\,$K.
Yet, a full transition was not observed.
With further reduction of the sample, the superconducting signature vanished completely, resulting in a dominantly IL-r phase thin film with more pronounced semiconducting/insulating transport properties (not shown here).

All results reported in sec.~\ref{subsubsec:annealtime} concerning vacuum annealing of SLCO on BTO/STO could be verified qualitatively for SLCO on KTO, too.
For simplicity, we will therefore only give a summary of the most important data and point out the main differences to SLCO on BTO/STO.
A series with vacuum annealing time varying between 5 and 20\,min showed, that $T_c$ was highest ($T_c = 18.3\,K$) when the room temperature resistivity was lowest ($\rho_\mathrm{300K} = 0.16\,$m$\Omega$cm).
However, we found a remarkable difference to SLCO on BTO/STO concerning the optimum annealing time.
Whereas SLCO on BTO/STO had an optimum annealing time $t_a^\mathrm{opt} = 22\,$min, SLCO on KTO required an annealing time of $t_a^\mathrm{opt} = 10\,$min, which corresponds to the value reported by Karimoto \textit{et al.}\cite{Karimoto01,Karimoto02}.
As discussed in sec.~\ref{subsec:charslcokto}, SLCO films on BTO/STO exhibit less tensile strain, as compared to the films grown on KTaO$_3$ substrates.
Karimoto \textit{et al.}\cite{Karimoto01} proposed that it is difficult to remove excess oxygen from compressively strained SLCO films.
This is due to a reduced in-plane lattice constant $a_\mathrm{SLCO}$, hindering the large O$^{2-}$ ions to diffuse and desorb from the crystal.
Regarding SLCO films on BTO/STO, this implies that longer annealing time is necessary to obtain a comparable degree of reduction as in the case of SLCO on KTO, conforming to our observation.
We further want to note, that RHEED patterns of SLCO on BTO/STO exhibited intermediate streaks (cf.~Fig.~\ref{fig:slcobtorheedafm}~(a)), pointing to a superstructure, possibly due to excess oxygen, which would well support the abovementioned interpretation.
\\
For SLCO on KTO, the lattice constants and the unit cell volume showed the same monotonic behavior as found for SLCO on BTO/STO, i.e.~with increasing vacuum annealing time, the $a$-axis increased, the $c$-axis decreased and the unit cell volume $V$ increased.
\\
Finally, too long vacuum annealing of SLCO on KTO ended up in the formation of an IL-r phase, which showed semiconducting electric transport behavior and was visible in optical micrographs.

\subsubsection{Vacuum annealing temperature}

\begin{figure*}[tb]
\centering
\ifgraph\includegraphics[width=1.00\textwidth]{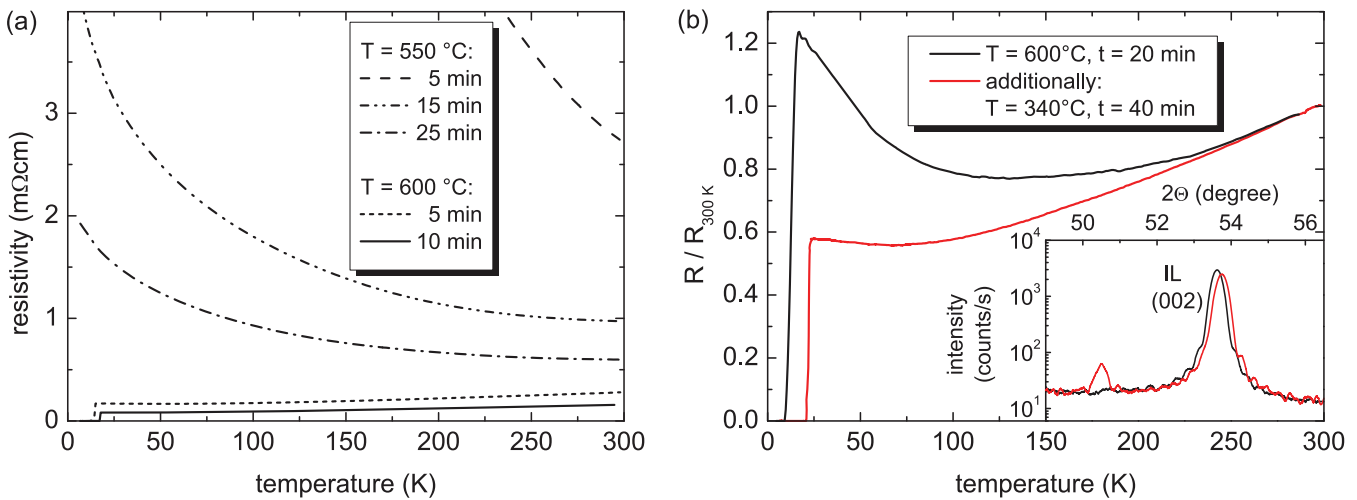}\fi
\caption{(Color online)
(a) Resistivity vs temperature for SLCO on KTO vacuum annealed under different conditions. The films of the first series were annealed at $T_a = 550\,^\circ$C for 5, 15, and 25\,min, and the films of the second set were annealed at $T_a = 600\,^\circ$C for 5 and 10\,min. (b) Normalized resistance vs temperature of an SLCO film on BTO/STO after 20\,min vacuum annealing at $T_a = 600\,^\circ$C and after additional 40\,min vacuum annealing at $T_a = 340\,^\circ$C. The inset shows XRD $\Theta - 2 \Theta$ scans before and after the second annealing step, yielding a $c$-axis constant of 3.415 and 3.407\,\AA, respectively. A small contribution of the IL-r phase is visible at $2 \Theta \approx 50.5\,^\circ$, after the additional vacuum annealing.
}
\label{fig:annealtemp}
\end{figure*}
%
In this section, we analyze the influence of vacuum annealing temperature $T_a$ on the properties of SLCO films.
We prepared two sets of SLCO films on KTO that were annealed at $T_a = 550\,^\circ$C and at $T_a = 600\,^\circ$C, respectively.
Figure \ref{fig:annealtemp}~(a) shows the $\rho(T)$ behavior of those films.
Regarding $T_a = 550\,^\circ$C, it is obvious that an increase of $t_a$ from 5 to 25\,min leads to a decrease of $\rho(T)$, in accordance with the results presented in sec.~\ref{subsubsec:annealtime}, which was explained by removal of excess oxygen.
However, to induce a superconducting transition, $t_a \gg 25\,$min is required at this reduced annealing temperature.
For comparison, two films annealed for $t_a = 5\,$min and 10\,min at $T_a = 600\,^\circ$C are additionally plotted in Fig.~\ref{fig:annealtemp}~(a).
Already after $t_a = 5\,$min vacuum annealing a superconducting transition with $T_c = 14.5\,$K was observed, which reached its maximum value of $T_c = 17.5\,$K after $t_a = 10\,$min.
In conclusion, removal of excess oxygen is strongly dependent on $T_a$, in accordance with literature\cite{Karimoto01,Li09}.
\\
For SLCO on BTO/STO we observed the qualitatively same behavior (not shown here), yet with double vacuum annealing time necessary, as described in sec.~\ref{subsubsec:annealtime}.

Finally, we want to describe a combination of high and low temperature vacuum annealing as introduced by Li \textit{et al.}\cite{Li09}.
As found by XRD and electric transport measurements, some films still contained too much excess oxygen, although having been vacuum annealed at $T_a = 600\,^\circ$C.
To remove the residual excess oxygen, those samples were additionally annealed in vacuum ($p_\mathrm{vac} \,\lapprox\, 10^{-5}\,$Pa) at $T_{a} \approx 340\,^\circ$C for $t_a = 40\,$min.
In most cases, the second low temperature vacuum annealing step was sufficient to enhance or induce superconductivity, as shown in Fig.~\ref{fig:annealtemp}~(b).
This can be explained by the fact that excess oxygen only occupies weakly bond interstitial sites and can thus diffuse and desorb even at low $T_a$.
Yet, it requires further examination whether \textit{all} excess oxygen can be removed at low $T_a$ or if a (previous) annealing step at high $T_a$ is crucial.
Furthermore, it was not essential if the low $T_a$ annealing step was performed after the sample had been exposed to ex-situ conditions or if it was performed in-situ directly after high-$T_a$ annealing;
both methods lead to comparable results.

\section{Conclusions}
\label{sec:Conclusions}

To summarize, we report in detail on the fabrication of single-crystalline thin films of the electron-doped infinite-layer superconductor Sr$_{1-x}$La$_x$CuO$_2$ (SLCO) by means of pulsed laser deposition and on their characterization by in-situ and ex-situ techniques.
[001]-oriented SrTiO$_3$ (STO) and KTaO$_3$ (KTO) single crystals were used as substrates.
Prior to deposition of SLCO on STO, a single-crystalline BaTiO$_3$ (BTO) thin film was deposited, acting as a buffer layer.
In case of KTO, no buffer layers were deposited.
The growth mode, the evolution of the in-plane lattice constant $a$, and the morphology of the BTO and SLCO films were monitored in-situ by high-pressure reflection high-energy electron diffraction.
We observed a Stranski-Krastanov growth mode, both for BTO and SLCO films and found that BTO films relaxed after growth of a few unit cells.
Atomic force microscopy revealed very flat surfaces of BTO and SLCO with asperities in the range of a few unit cells.
X-ray diffraction was used to determine the lattice constants of the films.
Fringes and narrow rocking curves indicated high crystalline quality.
A comparison of our thin film lattice constants $a$ and $c$ with literature bulk values showed that BTO buffer layers were (almost) relaxed, SLCO films on BTO-buffered STO (BTO/STO) moderately tensile strained, and SLCO films on KTO highly tensile strained.
However, SLCO films on both kinds of substrates showed comparable electric transport properties.
Furthermore, the stoichiometry of SLCO films on BTO/STO was determined by Rutherford backscattering spectroscopy.
It revealed slight overdoping ($x \approx 0.15$) as compared to optimally doped SLCO polycrystals ($x = 0.10$).
Moreover, we determined a critical current density of $j_c^\mathrm{4.2K} \approx 2 \times 10^6\,$A/cm$^2$ and a maximum $T_c$ of $\sim 22\,\mathrm{K} \approx \nicefrac{1}{2} T_c^\mathrm{bulk}$.
Finally, we discussed the influence of various process parameters on the thin film properties, such as varying excimer laser energy $E_L$, target-to-substrate
distance $d_\mathrm{TS}$, deposition pressure $p_\mathrm{O_2}$, vacuum annealing time $t_a$, and vacuum annealing temperature $T_a$ and show that these parameters directly influence the oxygen content of the SLCO films and hence their electric transport properties.

\acknowledgments
{
J.~T. gratefully acknowledges support by the Evangelisches Studienwerk e.V. Villigst.
V.~L. acknowledges partial financial support by a grant of the Romanian National Authority for Scientific Research, CNCS – UEFISCDI, project number PN-II-ID-PCE-2011-3-1065.
The authors thank Marcel Kimmerle for technical assistance with RBS measurements.
This work was funded by the Deutsche Forschungsgemeinschaft (project KL 930/11)
}.

\bibliography{SLCO_bibliography}

\begin{thebibliography}{53}
\expandafter\ifx\csname natexlab\endcsname\relax\def\natexlab#1{#1}\fi
\expandafter\ifx\csname bibnamefont\endcsname\relax
  \def\bibnamefont#1{#1}\fi
\expandafter\ifx\csname bibfnamefont\endcsname\relax
  \def\bibfnamefont#1{#1}\fi
\expandafter\ifx\csname citenamefont\endcsname\relax
  \def\citenamefont#1{#1}\fi
\expandafter\ifx\csname url\endcsname\relax
  \def\url#1{\texttt{#1}}\fi
\expandafter\ifx\csname urlprefix\endcsname\relax\def\urlprefix{URL }\fi
\providecommand{\bibinfo}[2]{#2}
\providecommand{\eprint}[2][]{\url{#2}}

\bibitem[{\citenamefont{Tokura et~al.}(1989)\citenamefont{Tokura, Takagi, and
  Uchida}}]{Tokura89}
\bibinfo{author}{\bibfnamefont{Y.}~\bibnamefont{Tokura}},
  \bibinfo{author}{\bibfnamefont{H.}~\bibnamefont{Takagi}}, \bibnamefont{and}
  \bibinfo{author}{\bibfnamefont{S.}~\bibnamefont{Uchida}},
  \bibinfo{journal}{Nature} \textbf{\bibinfo{volume}{337}},
  \bibinfo{pages}{345} (\bibinfo{year}{1989}).

\bibitem[{\citenamefont{Takagi et~al.}(1989)\citenamefont{Takagi, Uchida, and
  Tokura}}]{Takagi89}
\bibinfo{author}{\bibfnamefont{H.}~\bibnamefont{Takagi}},
  \bibinfo{author}{\bibfnamefont{S.}~\bibnamefont{Uchida}}, \bibnamefont{and}
  \bibinfo{author}{\bibfnamefont{Y.}~\bibnamefont{Tokura}},
  \bibinfo{journal}{Phys. Rev. Lett.} \textbf{\bibinfo{volume}{62}},
  \bibinfo{pages}{1197} (\bibinfo{year}{1989}).

\bibitem[{\citenamefont{Yamada et~al.}(1994)\citenamefont{Yamada, Kinoshita,
  and Shibata}}]{Yamada94}
\bibinfo{author}{\bibfnamefont{T.}~\bibnamefont{Yamada}},
  \bibinfo{author}{\bibfnamefont{K.}~\bibnamefont{Kinoshita}},
  \bibnamefont{and} \bibinfo{author}{\bibfnamefont{H.}~\bibnamefont{Shibata}},
  \bibinfo{journal}{Jpn. J. Appl. Phys.} \textbf{\bibinfo{volume}{33}},
  \bibinfo{pages}{L168} (\bibinfo{year}{1994}).

\bibitem[{\citenamefont{Naito and Hepp}(2000)}]{Naito00}
\bibinfo{author}{\bibfnamefont{M.}~\bibnamefont{Naito}} \bibnamefont{and}
  \bibinfo{author}{\bibfnamefont{M.}~\bibnamefont{Hepp}},
  \bibinfo{journal}{Jpn. J. Appl. Phys.} \textbf{\bibinfo{volume}{39}},
  \bibinfo{pages}{L485} (\bibinfo{year}{2000}).

\bibitem[{\citenamefont{Siegrist et~al.}(1988)\citenamefont{Siegrist, Zahurak,
  Murphy, and Roth}}]{Siegrist88}
\bibinfo{author}{\bibfnamefont{T.}~\bibnamefont{Siegrist}},
  \bibinfo{author}{\bibfnamefont{S.~M.} \bibnamefont{Zahurak}},
  \bibinfo{author}{\bibfnamefont{D.~W.} \bibnamefont{Murphy}},
  \bibnamefont{and} \bibinfo{author}{\bibfnamefont{R.~S.} \bibnamefont{Roth}},
  \bibinfo{journal}{Nature} \textbf{\bibinfo{volume}{334}},
  \bibinfo{pages}{231} (\bibinfo{year}{1988}).

\bibitem[{\citenamefont{Smith et~al.}(1991)\citenamefont{Smith, Manthiram,
  Zhou, Goodenough, and Markert}}]{Smith91}
\bibinfo{author}{\bibfnamefont{M.~G.} \bibnamefont{Smith}},
  \bibinfo{author}{\bibfnamefont{A.}~\bibnamefont{Manthiram}},
  \bibinfo{author}{\bibfnamefont{J.}~\bibnamefont{Zhou}},
  \bibinfo{author}{\bibfnamefont{J.~B.} \bibnamefont{Goodenough}},
  \bibnamefont{and} \bibinfo{author}{\bibfnamefont{J.~T.}
  \bibnamefont{Markert}}, \bibinfo{journal}{Nature}
  \textbf{\bibinfo{volume}{351}}, \bibinfo{pages}{549} (\bibinfo{year}{1991}).

\bibitem[{\citenamefont{Er et~al.}(1991)\citenamefont{Er, Miyamoto, Kanamaru,
  and Kikkawa}}]{Er91}
\bibinfo{author}{\bibfnamefont{G.}~\bibnamefont{Er}},
  \bibinfo{author}{\bibfnamefont{Y.}~\bibnamefont{Miyamoto}},
  \bibinfo{author}{\bibfnamefont{F.}~\bibnamefont{Kanamaru}}, \bibnamefont{and}
  \bibinfo{author}{\bibfnamefont{S.}~\bibnamefont{Kikkawa}},
  \bibinfo{journal}{Physica C} \textbf{\bibinfo{volume}{181}},
  \bibinfo{pages}{206} (\bibinfo{year}{1991}).

\bibitem[{\citenamefont{Er et~al.}(1992)\citenamefont{Er, Kikkawa, Kanamaru,
  Miyamoto, Tanaka, Sera, Sato, Hiroi, Takano, and Bando}}]{Er92}
\bibinfo{author}{\bibfnamefont{G.}~\bibnamefont{Er}},
  \bibinfo{author}{\bibfnamefont{S.}~\bibnamefont{Kikkawa}},
  \bibinfo{author}{\bibfnamefont{F.}~\bibnamefont{Kanamaru}},
  \bibinfo{author}{\bibfnamefont{Y.}~\bibnamefont{Miyamoto}},
  \bibinfo{author}{\bibfnamefont{S.}~\bibnamefont{Tanaka}},
  \bibinfo{author}{\bibfnamefont{M.}~\bibnamefont{Sera}},
  \bibinfo{author}{\bibfnamefont{M.}~\bibnamefont{Sato}},
  \bibinfo{author}{\bibfnamefont{Z.}~\bibnamefont{Hiroi}},
  \bibinfo{author}{\bibfnamefont{M.}~\bibnamefont{Takano}}, \bibnamefont{and}
  \bibinfo{author}{\bibfnamefont{Y.}~\bibnamefont{Bando}},
  \bibinfo{journal}{Physica C} \textbf{\bibinfo{volume}{196}},
  \bibinfo{pages}{271} (\bibinfo{year}{1992}).

\bibitem[{\citenamefont{Jorgensen et~al.}(1993)\citenamefont{Jorgensen,
  Radaelli, Hinks, Wagner, Kikkawa, Er, and Kanamaru}}]{Jorgensen93}
\bibinfo{author}{\bibfnamefont{J.~D.} \bibnamefont{Jorgensen}},
  \bibinfo{author}{\bibfnamefont{P.~G.} \bibnamefont{Radaelli}},
  \bibinfo{author}{\bibfnamefont{D.~G.} \bibnamefont{Hinks}},
  \bibinfo{author}{\bibfnamefont{J.~L.} \bibnamefont{Wagner}},
  \bibinfo{author}{\bibfnamefont{S.}~\bibnamefont{Kikkawa}},
  \bibinfo{author}{\bibfnamefont{G.}~\bibnamefont{Er}}, \bibnamefont{and}
  \bibinfo{author}{\bibfnamefont{F.}~\bibnamefont{Kanamaru}},
  \bibinfo{journal}{Phys. Rev. B} \textbf{\bibinfo{volume}{47}},
  \bibinfo{pages}{14654} (\bibinfo{year}{1993}).

\bibitem[{\citenamefont{Ikeda et~al.}(1993)\citenamefont{Ikeda, Hiroi, Azuma,
  Takano, Bando, and Takeda}}]{Ikeda93}
\bibinfo{author}{\bibfnamefont{N.}~\bibnamefont{Ikeda}},
  \bibinfo{author}{\bibfnamefont{Z.}~\bibnamefont{Hiroi}},
  \bibinfo{author}{\bibfnamefont{M.}~\bibnamefont{Azuma}},
  \bibinfo{author}{\bibfnamefont{M.}~\bibnamefont{Takano}},
  \bibinfo{author}{\bibfnamefont{Y.}~\bibnamefont{Bando}}, \bibnamefont{and}
  \bibinfo{author}{\bibfnamefont{Y.}~\bibnamefont{Takeda}},
  \bibinfo{journal}{Physica C} \textbf{\bibinfo{volume}{210}},
  \bibinfo{pages}{367} (\bibinfo{year}{1993}).

\bibitem[{\citenamefont{Torardi et~al.}(1988)\citenamefont{Torardi,
  Subramanian, Calabrese, Gopalakrishnan, Morrissey, Askew, Flippen, Chowdhry,
  and Sleight}}]{Torardi88}
\bibinfo{author}{\bibfnamefont{C.~C.} \bibnamefont{Torardi}},
  \bibinfo{author}{\bibfnamefont{M.~A.} \bibnamefont{Subramanian}},
  \bibinfo{author}{\bibfnamefont{J.~C.} \bibnamefont{Calabrese}},
  \bibinfo{author}{\bibfnamefont{J.}~\bibnamefont{Gopalakrishnan}},
  \bibinfo{author}{\bibfnamefont{K.~J.} \bibnamefont{Morrissey}},
  \bibinfo{author}{\bibfnamefont{T.~R.} \bibnamefont{Askew}},
  \bibinfo{author}{\bibfnamefont{R.~B.} \bibnamefont{Flippen}},
  \bibinfo{author}{\bibfnamefont{U.}~\bibnamefont{Chowdhry}}, \bibnamefont{and}
  \bibinfo{author}{\bibfnamefont{A.~W.} \bibnamefont{Sleight}},
  \bibinfo{journal}{Science} \textbf{\bibinfo{volume}{240}},
  \bibinfo{pages}{631} (\bibinfo{year}{1988}).

\bibitem[{\citenamefont{Shaked et~al.}(1995)\citenamefont{Shaked, Shimakawa,
  Hunter, Hitterman, Jorgensen, Han, and Payne}}]{Shaked95}
\bibinfo{author}{\bibfnamefont{H.}~\bibnamefont{Shaked}},
  \bibinfo{author}{\bibfnamefont{Y.}~\bibnamefont{Shimakawa}},
  \bibinfo{author}{\bibfnamefont{B.~A.} \bibnamefont{Hunter}},
  \bibinfo{author}{\bibfnamefont{R.~L.} \bibnamefont{Hitterman}},
  \bibinfo{author}{\bibfnamefont{J.~D.} \bibnamefont{Jorgensen}},
  \bibinfo{author}{\bibfnamefont{P.~D.} \bibnamefont{Han}}, \bibnamefont{and}
  \bibinfo{author}{\bibfnamefont{D.~A.} \bibnamefont{Payne}},
  \bibinfo{journal}{Phys. Rev. B} \textbf{\bibinfo{volume}{51}},
  \bibinfo{pages}{11784} (\bibinfo{year}{1995}).

\bibitem[{\citenamefont{Li et~al.}(1992)\citenamefont{Li, Kanai, Kawai, and
  Kawai}}]{Li92}
\bibinfo{author}{\bibfnamefont{X.}~\bibnamefont{Li}},
  \bibinfo{author}{\bibfnamefont{M.}~\bibnamefont{Kanai}},
  \bibinfo{author}{\bibfnamefont{T.}~\bibnamefont{Kawai}}, \bibnamefont{and}
  \bibinfo{author}{\bibfnamefont{S.}~\bibnamefont{Kawai}},
  \bibinfo{journal}{Jpn. J. Appl. Phys.} \textbf{\bibinfo{volume}{31}},
  \bibinfo{pages}{L217} (\bibinfo{year}{1992}).

\bibitem[{\citenamefont{Terashima et~al.}(1993)\citenamefont{Terashima, Sato,
  Takeno, Nakamura, and Miura}}]{Terashima93}
\bibinfo{author}{\bibfnamefont{Y.}~\bibnamefont{Terashima}},
  \bibinfo{author}{\bibfnamefont{R.}~\bibnamefont{Sato}},
  \bibinfo{author}{\bibfnamefont{S.}~\bibnamefont{Takeno}},
  \bibinfo{author}{\bibfnamefont{S.-I.} \bibnamefont{Nakamura}},
  \bibnamefont{and} \bibinfo{author}{\bibfnamefont{T.}~\bibnamefont{Miura}},
  \bibinfo{journal}{Jpn. J. Appl. Phys.} \textbf{\bibinfo{volume}{32}},
  \bibinfo{pages}{L48} (\bibinfo{year}{1993}).

\bibitem[{\citenamefont{Niu and Lieber}(1992)}]{Niu92}
\bibinfo{author}{\bibfnamefont{C.}~\bibnamefont{Niu}} \bibnamefont{and}
  \bibinfo{author}{\bibfnamefont{C.~M.} \bibnamefont{Lieber}},
  \bibinfo{journal}{Appl. Phys. Lett.} \textbf{\bibinfo{volume}{61}},
  \bibinfo{pages}{1712} (\bibinfo{year}{1992}).

\bibitem[{\citenamefont{Sugii et~al.}(1992)\citenamefont{Sugii, Ichikawa, Kubo,
  Sakurai, Yamamoto, and Yamauchi}}]{Sugii92}
\bibinfo{author}{\bibfnamefont{N.}~\bibnamefont{Sugii}},
  \bibinfo{author}{\bibfnamefont{M.}~\bibnamefont{Ichikawa}},
  \bibinfo{author}{\bibfnamefont{K.}~\bibnamefont{Kubo}},
  \bibinfo{author}{\bibfnamefont{T.}~\bibnamefont{Sakurai}},
  \bibinfo{author}{\bibfnamefont{K.}~\bibnamefont{Yamamoto}}, \bibnamefont{and}
  \bibinfo{author}{\bibfnamefont{H.}~\bibnamefont{Yamauchi}},
  \bibinfo{journal}{Physica C} \textbf{\bibinfo{volume}{196}},
  \bibinfo{pages}{129} (\bibinfo{year}{1992}).

\bibitem[{\citenamefont{Sugii et~al.}(1993)\citenamefont{Sugii, Matsuura, Kubo,
  Yamamoto, and Ichikawa}}]{Sugii93}
\bibinfo{author}{\bibfnamefont{N.}~\bibnamefont{Sugii}},
  \bibinfo{author}{\bibfnamefont{K.}~\bibnamefont{Matsuura}},
  \bibinfo{author}{\bibfnamefont{K.}~\bibnamefont{Kubo}},
  \bibinfo{author}{\bibfnamefont{K.}~\bibnamefont{Yamamoto}}, \bibnamefont{and}
  \bibinfo{author}{\bibfnamefont{M.}~\bibnamefont{Ichikawa}},
  \bibinfo{journal}{J. Appl. Phys.} \textbf{\bibinfo{volume}{74}},
  \bibinfo{pages}{4047} (\bibinfo{year}{1993}).

\bibitem[{\citenamefont{Markert et~al.}(2000)\citenamefont{Markert, Messina,
  Dam, Huijbregste, Rector, and Griessen}}]{Markert00}
\bibinfo{author}{\bibfnamefont{J.~T.} \bibnamefont{Markert}},
  \bibinfo{author}{\bibfnamefont{T.~C.} \bibnamefont{Messina}},
  \bibinfo{author}{\bibfnamefont{B.}~\bibnamefont{Dam}},
  \bibinfo{author}{\bibfnamefont{J.}~\bibnamefont{Huijbregste}},
  \bibinfo{author}{\bibfnamefont{J.~H.} \bibnamefont{Rector}},
  \bibnamefont{and} \bibinfo{author}{\bibfnamefont{R.}~\bibnamefont{Griessen}},
  \bibinfo{journal}{Proc. SPIE} \textbf{\bibinfo{volume}{4058}},
  \bibinfo{pages}{141} (\bibinfo{year}{2000}).

\bibitem[{\citenamefont{Markert et~al.}(2003)\citenamefont{Markert, Messina,
  Dam, Huijbregste, Rector, and Griessen}}]{Markert03}
\bibinfo{author}{\bibfnamefont{J.~T.} \bibnamefont{Markert}},
  \bibinfo{author}{\bibfnamefont{T.~C.} \bibnamefont{Messina}},
  \bibinfo{author}{\bibfnamefont{B.}~\bibnamefont{Dam}},
  \bibinfo{author}{\bibfnamefont{J.}~\bibnamefont{Huijbregste}},
  \bibinfo{author}{\bibfnamefont{J.~H.} \bibnamefont{Rector}},
  \bibnamefont{and} \bibinfo{author}{\bibfnamefont{R.}~\bibnamefont{Griessen}},
  \bibinfo{journal}{IEEE} \textbf{\bibinfo{volume}{13}}, \bibinfo{pages}{2684}
  (\bibinfo{year}{2003}).

\bibitem[{\citenamefont{Leca et~al.}(2006)\citenamefont{Leca, Blank, Rijnders,
  Bals, and van Tendeloo}}]{Leca06}
\bibinfo{author}{\bibfnamefont{V.}~\bibnamefont{Leca}},
  \bibinfo{author}{\bibfnamefont{D.~H.~A.} \bibnamefont{Blank}},
  \bibinfo{author}{\bibfnamefont{G.}~\bibnamefont{Rijnders}},
  \bibinfo{author}{\bibfnamefont{S.}~\bibnamefont{Bals}}, \bibnamefont{and}
  \bibinfo{author}{\bibfnamefont{G.}~\bibnamefont{van Tendeloo}},
  \bibinfo{journal}{Appl. Phys. Lett.} \textbf{\bibinfo{volume}{89}},
  \bibinfo{pages}{092504} (\bibinfo{year}{2006}).

\bibitem[{\citenamefont{Leca et~al.}(2008)\citenamefont{Leca, Visanescu, Back,
  Kleiner, and Koelle}}]{Leca08}
\bibinfo{author}{\bibfnamefont{V.}~\bibnamefont{Leca}},
  \bibinfo{author}{\bibfnamefont{G.}~\bibnamefont{Visanescu}},
  \bibinfo{author}{\bibfnamefont{C.}~\bibnamefont{Back}},
  \bibinfo{author}{\bibfnamefont{R.}~\bibnamefont{Kleiner}}, \bibnamefont{and}
  \bibinfo{author}{\bibfnamefont{D.}~\bibnamefont{Koelle}},
  \bibinfo{journal}{Appl. Phys. A} \textbf{\bibinfo{volume}{93}},
  \bibinfo{pages}{779} (\bibinfo{year}{2008}).

\bibitem[{\citenamefont{Karimoto et~al.}(2001)\citenamefont{Karimoto, Ueda,
  Naito, and Imai}}]{Karimoto01}
\bibinfo{author}{\bibfnamefont{S.}~\bibnamefont{Karimoto}},
  \bibinfo{author}{\bibfnamefont{K.}~\bibnamefont{Ueda}},
  \bibinfo{author}{\bibfnamefont{M.}~\bibnamefont{Naito}}, \bibnamefont{and}
  \bibinfo{author}{\bibfnamefont{T.}~\bibnamefont{Imai}},
  \bibinfo{journal}{Appl. Phys. Lett.} \textbf{\bibinfo{volume}{79}},
  \bibinfo{pages}{2767} (\bibinfo{year}{2001}).

\bibitem[{\citenamefont{Karimoto and Naito}(2004{\natexlab{a}})}]{Karimoto04}
\bibinfo{author}{\bibfnamefont{S.-I.} \bibnamefont{Karimoto}} \bibnamefont{and}
  \bibinfo{author}{\bibfnamefont{M.}~\bibnamefont{Naito}},
  \bibinfo{journal}{Appl. Phys. Lett.} \textbf{\bibinfo{volume}{84}},
  \bibinfo{pages}{2136} (\bibinfo{year}{2004}{\natexlab{a}}).

\bibitem[{\citenamefont{Jorgensen et~al.}(1994)\citenamefont{Jorgensen,
  Radaelli, Shaked, Wagner, Hunter, Mitchell, Hitterman, and
  Hinks}}]{Jorgensen94}
\bibinfo{author}{\bibfnamefont{J.~D.} \bibnamefont{Jorgensen}},
  \bibinfo{author}{\bibfnamefont{P.~G.} \bibnamefont{Radaelli}},
  \bibinfo{author}{\bibfnamefont{H.}~\bibnamefont{Shaked}},
  \bibinfo{author}{\bibfnamefont{J.~L.} \bibnamefont{Wagner}},
  \bibinfo{author}{\bibfnamefont{B.~A.} \bibnamefont{Hunter}},
  \bibinfo{author}{\bibfnamefont{J.~F.} \bibnamefont{Mitchell}},
  \bibinfo{author}{\bibfnamefont{R.~L.} \bibnamefont{Hitterman}},
  \bibnamefont{and} \bibinfo{author}{\bibfnamefont{D.~G.} \bibnamefont{Hinks}},
  \bibinfo{journal}{Journal of Superconductivity} \textbf{\bibinfo{volume}{7}},
  \bibinfo{pages}{145} (\bibinfo{year}{1994}).

\bibitem[{\citenamefont{Naito et~al.}(1997)\citenamefont{Naito, Sato, and
  Yamamoto}}]{Naito97}
\bibinfo{author}{\bibfnamefont{M.}~\bibnamefont{Naito}},
  \bibinfo{author}{\bibfnamefont{H.}~\bibnamefont{Sato}}, \bibnamefont{and}
  \bibinfo{author}{\bibfnamefont{H.}~\bibnamefont{Yamamoto}},
  \bibinfo{journal}{Physica C} \textbf{\bibinfo{volume}{293}},
  \bibinfo{pages}{36} (\bibinfo{year}{1997}).

\bibitem[{\citenamefont{Adachi et~al.}(1992)\citenamefont{Adachi, Satoh,
  Ichikawa, Setsune, and Wasa}}]{Adachi92}
\bibinfo{author}{\bibfnamefont{H.}~\bibnamefont{Adachi}},
  \bibinfo{author}{\bibfnamefont{T.}~\bibnamefont{Satoh}},
  \bibinfo{author}{\bibfnamefont{Y.}~\bibnamefont{Ichikawa}},
  \bibinfo{author}{\bibfnamefont{K.}~\bibnamefont{Setsune}}, \bibnamefont{and}
  \bibinfo{author}{\bibfnamefont{K.}~\bibnamefont{Wasa}},
  \bibinfo{journal}{Physica C} \textbf{\bibinfo{volume}{196}},
  \bibinfo{pages}{14} (\bibinfo{year}{1992}).

\bibitem[{\citenamefont{Li et~al.}(2009)\citenamefont{Li, Jovanovic, Raffy, and
  Megtert}}]{Li09}
\bibinfo{author}{\bibfnamefont{Z.~Z.} \bibnamefont{Li}},
  \bibinfo{author}{\bibfnamefont{V.}~\bibnamefont{Jovanovic}},
  \bibinfo{author}{\bibfnamefont{H.}~\bibnamefont{Raffy}}, \bibnamefont{and}
  \bibinfo{author}{\bibfnamefont{S.}~\bibnamefont{Megtert}},
  \bibinfo{journal}{Physica C} \textbf{\bibinfo{volume}{469}},
  \bibinfo{pages}{73} (\bibinfo{year}{2009}).

\bibitem[{\citenamefont{Bals et~al.}(2003)\citenamefont{Bals, Tendeloo,
  Rijnders, Huijben, Leca, and Blank}}]{Bals03}
\bibinfo{author}{\bibfnamefont{S.}~\bibnamefont{Bals}},
  \bibinfo{author}{\bibfnamefont{G.~V.} \bibnamefont{Tendeloo}},
  \bibinfo{author}{\bibfnamefont{G.}~\bibnamefont{Rijnders}},
  \bibinfo{author}{\bibfnamefont{M.}~\bibnamefont{Huijben}},
  \bibinfo{author}{\bibfnamefont{V.}~\bibnamefont{Leca}}, \bibnamefont{and}
  \bibinfo{author}{\bibfnamefont{D.~H.~A.} \bibnamefont{Blank}},
  \bibinfo{journal}{IEEE Trans. Appl. Supercond.}
  \textbf{\bibinfo{volume}{13}}, \bibinfo{pages}{2834} (\bibinfo{year}{2003}).

\bibitem[{\citenamefont{Zhou et~al.}(1993)\citenamefont{Zhou, Yao, Li, Xu, Jia,
  and Zhao}}]{Zhou93}
\bibinfo{author}{\bibfnamefont{X.}~\bibnamefont{Zhou}},
  \bibinfo{author}{\bibfnamefont{Y.}~\bibnamefont{Yao}},
  \bibinfo{author}{\bibfnamefont{J.}~\bibnamefont{Li}},
  \bibinfo{author}{\bibfnamefont{W.}~\bibnamefont{Xu}},
  \bibinfo{author}{\bibfnamefont{S.}~\bibnamefont{Jia}}, \bibnamefont{and}
  \bibinfo{author}{\bibfnamefont{Z.}~\bibnamefont{Zhao}},
  \bibinfo{journal}{Chinese. Phys. Lett.} \textbf{\bibinfo{volume}{10}},
  \bibinfo{pages}{503} (\bibinfo{year}{1993}).

\bibitem[{\citenamefont{Mercey et~al.}(1995)\citenamefont{Mercey, Gupta,
  Hervieu, and Raveau}}]{Mercey95}
\bibinfo{author}{\bibfnamefont{B.}~\bibnamefont{Mercey}},
  \bibinfo{author}{\bibfnamefont{A.}~\bibnamefont{Gupta}},
  \bibinfo{author}{\bibfnamefont{M.}~\bibnamefont{Hervieu}}, \bibnamefont{and}
  \bibinfo{author}{\bibfnamefont{B.}~\bibnamefont{Raveau}},
  \bibinfo{journal}{Journal of Solid State Chemistry}
  \textbf{\bibinfo{volume}{116}}, \bibinfo{pages}{300} (\bibinfo{year}{1995}).

\bibitem[{\citenamefont{Leca}(2003)}]{Leca03}
\bibinfo{author}{\bibfnamefont{V.}~\bibnamefont{Leca}}, Ph.D. thesis,
  \bibinfo{school}{University of Twente, Enschede, The Netherlands}
  (\bibinfo{year}{2003}), \bibinfo{note}{iSBN 90-365-1928-4}.

\bibitem[{Che()}]{Chemco}
\bibinfo{note}{Chemco GmbH, Germany}.

\bibitem[{Cry({\natexlab{a}})}]{Crystal}
\bibinfo{note}{Crystal GmbH, Germany}.

\bibitem[{Cry({\natexlab{b}})}]{Crystec}
\bibinfo{note}{CrysTec GmbH, Germany}.

\bibitem[{Ros()}]{Rosenau}
\bibinfo{note}{Rosenau accelerator, Physikalisches Institut, Universit\"{a}t
  T\"{u}bingen, Germany}.

\bibitem[{\citenamefont{Diebold}(2010)}]{Diebold10}
\bibinfo{author}{\bibfnamefont{S.}~\bibnamefont{Diebold}}, Master's thesis,
  \bibinfo{school}{Universit\"{a}t T\"{u}bingen, Germany}
  (\bibinfo{year}{2010}).

\bibitem[{CRC()}]{CRC71}
\bibinfo{note}{Handbook of Chemistry and Physics, The Chemical Rubber Company,
  U. S. A., 52nd ed. (1971)}.

\bibitem[{\citenamefont{Dungan et~al.}(1952)\citenamefont{Dungan, Kane, and
  L.~R.~Bickford}}]{Dungan52}
\bibinfo{author}{\bibfnamefont{R.~H.} \bibnamefont{Dungan}},
  \bibinfo{author}{\bibfnamefont{D.~F.} \bibnamefont{Kane}}, \bibnamefont{and}
  \bibinfo{author}{\bibfnamefont{J.}~\bibnamefont{L.~R.~Bickford}},
  \bibinfo{journal}{J. Am. Cer. Soc.} \textbf{\bibinfo{volume}{35}},
  \bibinfo{pages}{318} (\bibinfo{year}{1952}).

\bibitem[{\citenamefont{Donohue et~al.}(1958)\citenamefont{Donohue, Miller, and
  Cline}}]{Donohue58}
\bibinfo{author}{\bibfnamefont{J.}~\bibnamefont{Donohue}},
  \bibinfo{author}{\bibfnamefont{S.~J.} \bibnamefont{Miller}},
  \bibnamefont{and} \bibinfo{author}{\bibfnamefont{R.~F.} \bibnamefont{Cline}},
  \bibinfo{journal}{Acta Cryst.} \textbf{\bibinfo{volume}{11}},
  \bibinfo{pages}{693} (\bibinfo{year}{1958}).

\bibitem[{\citenamefont{Stranski and Krastanov}(1939)}]{Stranski39}
\bibinfo{author}{\bibfnamefont{I.~N.} \bibnamefont{Stranski}} \bibnamefont{and}
  \bibinfo{author}{\bibfnamefont{L.}~\bibnamefont{Krastanov}},
  \bibinfo{journal}{Akad. Wiss. Lit. Mainz Math.-Natur. Kl. IIb}
  \textbf{\bibinfo{volume}{146}}, \bibinfo{pages}{797} (\bibinfo{year}{1939}).

\bibitem[{\citenamefont{Werner et~al.}(2009)\citenamefont{Werner, Raisch, Leca,
  Ion, Bals, VanTendeloo, Chass\'{e}, Kleiner, and Koelle}}]{Werner09}
\bibinfo{author}{\bibfnamefont{R.}~\bibnamefont{Werner}},
  \bibinfo{author}{\bibfnamefont{C.}~\bibnamefont{Raisch}},
  \bibinfo{author}{\bibfnamefont{V.}~\bibnamefont{Leca}},
  \bibinfo{author}{\bibfnamefont{V.}~\bibnamefont{Ion}},
  \bibinfo{author}{\bibfnamefont{S.}~\bibnamefont{Bals}},
  \bibinfo{author}{\bibfnamefont{G.}~\bibnamefont{VanTendeloo}},
  \bibinfo{author}{\bibfnamefont{T.}~\bibnamefont{Chass\'{e}}},
  \bibinfo{author}{\bibfnamefont{R.}~\bibnamefont{Kleiner}}, \bibnamefont{and}
  \bibinfo{author}{\bibfnamefont{D.}~\bibnamefont{Koelle}},
  \bibinfo{journal}{Phys. Rev. B} \textbf{\bibinfo{volume}{79}},
  \bibinfo{pages}{054416} (\bibinfo{year}{2009}).

\bibitem[{\citenamefont{Warren}()}]{Warren69}
\bibinfo{author}{\bibfnamefont{B.~E.} \bibnamefont{Warren}},
  \bibinfo{note}{x-Ray Diffraction, Addison-Wesley (1991)}.

\bibitem[{\citenamefont{Terai et~al.}(2002)\citenamefont{Terai, Lippmaa, Ahmet,
  Chikyow, Fujii, Koinuma, and Kawasaki}}]{Terai02}
\bibinfo{author}{\bibfnamefont{K.}~\bibnamefont{Terai}},
  \bibinfo{author}{\bibfnamefont{M.}~\bibnamefont{Lippmaa}},
  \bibinfo{author}{\bibfnamefont{P.}~\bibnamefont{Ahmet}},
  \bibinfo{author}{\bibfnamefont{T.}~\bibnamefont{Chikyow}},
  \bibinfo{author}{\bibfnamefont{T.}~\bibnamefont{Fujii}},
  \bibinfo{author}{\bibfnamefont{H.}~\bibnamefont{Koinuma}}, \bibnamefont{and}
  \bibinfo{author}{\bibfnamefont{M.}~\bibnamefont{Kawasaki}},
  \bibinfo{journal}{Appl. Phys. Lett.} \textbf{\bibinfo{volume}{80}},
  \bibinfo{pages}{4437} (\bibinfo{year}{2002}).

\bibitem[{\citenamefont{Tomaschko et~al.}(2011)\citenamefont{Tomaschko, Raisch,
  Leca, Chass\'{e}, Kleiner, and Koelle}}]{Tomaschko11}
\bibinfo{author}{\bibfnamefont{J.}~\bibnamefont{Tomaschko}},
  \bibinfo{author}{\bibfnamefont{C.}~\bibnamefont{Raisch}},
  \bibinfo{author}{\bibfnamefont{V.}~\bibnamefont{Leca}},
  \bibinfo{author}{\bibfnamefont{T.}~\bibnamefont{Chass\'{e}}},
  \bibinfo{author}{\bibfnamefont{R.}~\bibnamefont{Kleiner}}, \bibnamefont{and}
  \bibinfo{author}{\bibfnamefont{D.}~\bibnamefont{Koelle}},
  \bibinfo{journal}{Phys. Rev. B} \textbf{\bibinfo{volume}{84}},
  \bibinfo{pages}{064521} (\bibinfo{year}{2011}).

\bibitem[{Sim()}]{Simnra}
\bibinfo{note}{SIMNRA, Max-Planck-Institut f\"{u}r Plasmaphysik, Garching,
  Germany}.

\bibitem[{\citenamefont{Ito et~al.}(1993)\citenamefont{Ito, Takenaka, and
  Uchida}}]{Ito93}
\bibinfo{author}{\bibfnamefont{T.}~\bibnamefont{Ito}},
  \bibinfo{author}{\bibfnamefont{K.}~\bibnamefont{Takenaka}}, \bibnamefont{and}
  \bibinfo{author}{\bibfnamefont{S.}~\bibnamefont{Uchida}},
  \bibinfo{journal}{Phys. Rev. Lett.} \textbf{\bibinfo{volume}{70}},
  \bibinfo{pages}{3995} (\bibinfo{year}{1993}).

\bibitem[{\citenamefont{Nishizaki et~al.}(1994)\citenamefont{Nishizaki,
  Yamasaki, Tanaka, Ichikawa, Fukami, Aomine, Kubo, and Suzuki}}]{Nishizaki94}
\bibinfo{author}{\bibfnamefont{T.}~\bibnamefont{Nishizaki}},
  \bibinfo{author}{\bibfnamefont{Y.}~\bibnamefont{Yamasaki}},
  \bibinfo{author}{\bibfnamefont{R.}~\bibnamefont{Tanaka}},
  \bibinfo{author}{\bibfnamefont{F.}~\bibnamefont{Ichikawa}},
  \bibinfo{author}{\bibfnamefont{T.}~\bibnamefont{Fukami}},
  \bibinfo{author}{\bibfnamefont{T.}~\bibnamefont{Aomine}},
  \bibinfo{author}{\bibfnamefont{S.}~\bibnamefont{Kubo}}, \bibnamefont{and}
  \bibinfo{author}{\bibfnamefont{M.}~\bibnamefont{Suzuki}},
  \bibinfo{journal}{Physica B} \textbf{\bibinfo{volume}{194-196}},
  \bibinfo{pages}{1877} (\bibinfo{year}{1994}).

\bibitem[{\citenamefont{Karimoto et~al.}(2002)\citenamefont{Karimoto, Ueda,
  Naito, and Imai}}]{Karimoto02}
\bibinfo{author}{\bibfnamefont{S.}~\bibnamefont{Karimoto}},
  \bibinfo{author}{\bibfnamefont{K.}~\bibnamefont{Ueda}},
  \bibinfo{author}{\bibfnamefont{M.}~\bibnamefont{Naito}}, \bibnamefont{and}
  \bibinfo{author}{\bibfnamefont{T.}~\bibnamefont{Imai}},
  \bibinfo{journal}{Physica C} \textbf{\bibinfo{volume}{378-381}},
  \bibinfo{pages}{127} (\bibinfo{year}{2002}).

\bibitem[{\citenamefont{Gaojie et~al.}(2001)\citenamefont{Gaojie, Qirong,
  Zengming, and Zejun}}]{Gaojie01}
\bibinfo{author}{\bibfnamefont{X.}~\bibnamefont{Gaojie}},
  \bibinfo{author}{\bibfnamefont{P.}~\bibnamefont{Qirong}},
  \bibinfo{author}{\bibfnamefont{Z.}~\bibnamefont{Zengming}}, \bibnamefont{and}
  \bibinfo{author}{\bibfnamefont{D.}~\bibnamefont{Zejun}},
  \bibinfo{journal}{Journal of Superconductivity: Incorporating Novel
  Magnetism} \textbf{\bibinfo{volume}{14}}, \bibinfo{pages}{509}
  (\bibinfo{year}{2001}).

\bibitem[{\citenamefont{Matsunaka et~al.}(2005)\citenamefont{Matsunaka,
  Rodulfo, and Kasai}}]{Matsunaka05}
\bibinfo{author}{\bibfnamefont{D.}~\bibnamefont{Matsunaka}},
  \bibinfo{author}{\bibfnamefont{E.~T.} \bibnamefont{Rodulfo}},
  \bibnamefont{and} \bibinfo{author}{\bibfnamefont{H.}~\bibnamefont{Kasai}},
  \bibinfo{journal}{Solid State Commun.} \textbf{\bibinfo{volume}{134}},
  \bibinfo{pages}{355} (\bibinfo{year}{2005}).

\bibitem[{\citenamefont{Nie et~al.}(2003{\natexlab{a}})\citenamefont{Nie,
  Badica, Hirai, Sundaresan, Crisan, Kit\^{o}, Terada, Iyo, Tanaka, and
  Ihara}}]{Nie03}
\bibinfo{author}{\bibfnamefont{J.~C.} \bibnamefont{Nie}},
  \bibinfo{author}{\bibfnamefont{P.}~\bibnamefont{Badica}},
  \bibinfo{author}{\bibfnamefont{M.}~\bibnamefont{Hirai}},
  \bibinfo{author}{\bibfnamefont{A.}~\bibnamefont{Sundaresan}},
  \bibinfo{author}{\bibfnamefont{A.}~\bibnamefont{Crisan}},
  \bibinfo{author}{\bibfnamefont{H.}~\bibnamefont{Kit\^{o}}},
  \bibinfo{author}{\bibfnamefont{N.}~\bibnamefont{Terada}},
  \bibinfo{author}{\bibfnamefont{A.}~\bibnamefont{Iyo}},
  \bibinfo{author}{\bibfnamefont{Y.}~\bibnamefont{Tanaka}}, \bibnamefont{and}
  \bibinfo{author}{\bibfnamefont{H.}~\bibnamefont{Ihara}},
  \bibinfo{journal}{Supercond. Sci. Technol.} \textbf{\bibinfo{volume}{16}},
  \bibinfo{pages}{L1} (\bibinfo{year}{2003}{\natexlab{a}}).

\bibitem[{\citenamefont{Nie et~al.}(2003{\natexlab{b}})\citenamefont{Nie,
  Badica, Hirai, Kodama, Crisan, Sundaresan, Tanaka, and Ihara}}]{Nie03a}
\bibinfo{author}{\bibfnamefont{J.~C.} \bibnamefont{Nie}},
  \bibinfo{author}{\bibfnamefont{P.}~\bibnamefont{Badica}},
  \bibinfo{author}{\bibfnamefont{M.}~\bibnamefont{Hirai}},
  \bibinfo{author}{\bibfnamefont{Y.}~\bibnamefont{Kodama}},
  \bibinfo{author}{\bibfnamefont{A.}~\bibnamefont{Crisan}},
  \bibinfo{author}{\bibfnamefont{A.}~\bibnamefont{Sundaresan}},
  \bibinfo{author}{\bibfnamefont{Y.}~\bibnamefont{Tanaka}}, \bibnamefont{and}
  \bibinfo{author}{\bibfnamefont{H.}~\bibnamefont{Ihara}},
  \bibinfo{journal}{Physica C} \textbf{\bibinfo{volume}{388-389}},
  \bibinfo{pages}{441} (\bibinfo{year}{2003}{\natexlab{b}}).

\bibitem[{\citenamefont{Karimoto and Naito}(2004{\natexlab{b}})}]{Karimoto04a}
\bibinfo{author}{\bibfnamefont{S.}~\bibnamefont{Karimoto}} \bibnamefont{and}
  \bibinfo{author}{\bibfnamefont{M.}~\bibnamefont{Naito}},
  \bibinfo{journal}{Physica C} \textbf{\bibinfo{volume}{412-414}},
  \bibinfo{pages}{1349} (\bibinfo{year}{2004}{\natexlab{b}}).

\end{thebibliography}

\end{document}